\documentclass[a4paper,11pt]{article}
\pdfoutput=1

\usepackage{amsmath,amssymb,bm,color}
\usepackage{appendix}
\usepackage{graphicx}
\usepackage{xcolor}
\usepackage{cancel}
\usepackage{jcappub}
\usepackage[T1]{fontenc} 
\usepackage{slashed}

\newcommand{\be}{\begin{equation}}
\newcommand{\ee}{\end{equation}}
\newcommand{\bea}{\begin{eqnarray}}
\newcommand{\eea}{\end{eqnarray}}

\title{The Primordial Black Holes that Disappeared: Connections to Dark Matter and MHz-GHz Gravitational Waves}

\author[a]{Thomas C. Gehrman,}

\author[b]{Barmak Shams Es Haghi,}

\author[a]{Kuver Sinha,}

\author[a]{and Tao Xu}

\affiliation[a]{Department of Physics and Astronomy, University of Oklahoma, Norman, OK 73019, USA}
\affiliation[b]{Texas Center for Cosmology and Astroparticle Physics, Weinberg Institute for Theoretical Physics, Department of Physics, The University of Texas at Austin, Austin, TX 78712, USA}

\emailAdd{thomas.gehrman@ou.edu}

\emailAdd{shams@austin.utexas.edu}

\emailAdd{kuver.sinha@ou.edu}

\emailAdd{tao.xu@ou.edu}

\abstract
{
In the post-LIGO era, there has been a lot of focus on primordial black holes (PBHs) heavier than $\sim 10^{15}$g as potential dark matter (DM) candidates. We point out that the branch of the PBH family that disappeared - PBHs lighter than $\sim 10^9$g that ostensibly Hawking evaporated away in the early Universe - also constitute an interesting frontier for DM physics. Hawking evaporation itself serves as a portal through which such PBHs can illuminate new physics, for example by  emitting dark sector particles. Taking a simple DM scalar singlet model as a template, we compute the abundance and mass of PBHs that could have provided, by Hawking evaporation,  the correct DM relic density. We consider two classes of such PBHs: those originating from curvature perturbations generated by inflation, and those originating from false vacuum collapse during a first-order phase transition. For PBHs of both origins we compute the  gravitational wave (GW) signals emanating from their formation stage: from second-order effects in the case of curvature perturbations, and from sound waves in the case of phase transitions. The 
GW signals have peak frequencies in the MHz-GHz range typical of such light PBHs. We compute the strength of such GWs compatible with the observed DM relic density, and find that the GW signal morphology can in principle allow one to distinguish between the two PBH formation histories. 
}

\begin{document}
\hfill{\small UTWI-10-2023}

\maketitle
\section{Introduction}
\label{sec:introduction}

A particularly illuminating way of classifying primordial black holes (PBHs) is the following: $(i)$ PBHs heavier than $\mathcal{O}(10^{15})g$ that are stable enough to have persisted to this day, and could constitute all or part of dark matter (DM); and $(ii)$ PBHs lighter than $\mathcal{O}(10^{9})g$ that disappeared via Hawking evaporation before Big Bang Nucleosynthesis (BBN). In the post-LIGO era, it is the first category that has garnered a lot of attention; justifiably so, since a variety of tests can be proposed to probe their existence. We refer to \cite{Caldwell:2022qsj, Bird:2016dcv, Clesse:2016vqa, Sasaki:2016jop, Kouvaris:2018wnh, Ali-Haimoud:2017rtz, Guo:2019sns, Guo:2017njn, Coogan:2020tuf, Clark:2018ghm, Laha:2020ivk,Carr:2009jm,Laha:2019ssq, Boudaud:2018hqb, Poulin:2016anj, Clark:2016nst, Khlopov:2008qy, Belotsky:2014kca, Belotsky:2018wph, Ketov:2019mfc, DeRocco:2019fjq, Kim:2020ngi, Saha:2021pqf,Green:2020jor, Kozaczuk:2021wcl, Mittal:2021egv, Agashe:2022jgk, Marfatia:2021hcp, Xie:2023cwi, Wang:2022nml} and references in the recent review \cite{Escriva:2022duf,Carr:2020gox}  as a sample of this vast literature.

In contrast, the second branch of the PBH family - those that disappeared early on - has received far less attention as far as their detection is concerned. The conventional requirement is that the Hawking evaporation should occur before BBN. More sophisticated bounds from BBN are reported in~\cite{Carr:2020gox}, using methods developed in~\cite{Kawasaki:2000en, Hasegawa:2019jsa} for constraining the reheating temperature from the PBH evaporation. In a sense, this branch of the PBH family shares the problem of all other putative relics that could have existed in the early Universe -- compatibility with BBN necessitates that they leave the Universe in a state of thermal equilibrium  by the time they vanish, obscuring their very existence, which must then be inferred indirectly. The example of cosmological moduli is instructive in this regard: their existence can be inferred from their effect on particle physics, for example the physics of DM, baryogenesis \cite{Kane:2015jia, Allahverdi:2020bys}, or cosmology~\cite{ShamsEsHaghi:2022azq}. This class of light PBHs are fascinating objects, serving as a portal to beyond-Standard Model physics. Since all dark/hidden sector particles couple at least to gravity, they would have been produced when PBHs underwent Hawking evaporation, providing a particularly rich and model-independent window into new physics\footnote{This can be contrasted with the case of cosmological moduli, whose coupling to hidden sector particles is much more model-dependent \cite{Allahverdi:2013noa}.}. Connections to DM \cite{Sandick:2021gew, Allahverdi:2017sks, Bell:1998jk, Lennon:2017tqq, Gondolo:2020uqv, Bernal:2021yyb, Cheek:2021odj, Cheek:2021cfe,Bhaumik:2022zdd}, dark radiation \cite{Arbey:2021ysg, Hooper:2019gtx, Masina:2021zpu,Cheek:2022dbx,Papanikolaou:2023oxq}, and baryogenesis \cite{Hawking:1975vcx, 1976ZhPmR..24...29Z, 1976ApJ...206....8C, PhysRevD.19.1036, Turner:1979bt, 1980PhLB...94..364G, Alexander:2007gj, Baumann:2007yr, Fujita:2014hha, Hook:2014mla,Hamada:2016jnq, Morrison:2018xla, Bernal:2022pue, Datta:2020bht, Bhaumik:2022pil, Gehrman:2022imk} in the primordial Universe, as well as axion-like particles at low redshift~\cite{Agashe:2022phd,Jho:2022wxd} have been investigated by various authors.

One could then ask: could the PBHs that disappeared early on have left behind tell-tale signatures of their existence? Here the answer can be in the affirmative depending on the PBH formation mechanism, although it requires us to probe an extremely challenging experimental frontier: ultra-high frequency (MHz-GHz) gravitational waves (GWs) \cite{Aggarwal:2020olq}. The reason is as follows. PBHs, during their formation stage, are typically accompanied by the emission of GWs. The precise origin of the GWs depends on the formation mechanism: for example, in the canonical example of PBHs coming from curvature perturbations during inflation, the source is second-order effects in perturbation theory \cite{1967PThPh..37..831T, Mollerach:2003nq, Ananda:2006af, Baumann:2007zm, Acquaviva:2002ud, Assadullahi:2009jc, Kohri:2018awv, Inomata:2018epa,Domenech:2021ztg,Inomata:2019ivs, Inomata:2016rbd, Espinosa:2018eve, Braglia:2020eai,Pi:2020otn, Yuan:2021qgz}; on the other hand, for the case of PBHs formed during a first-order phase transition (FOPT), the FOPT itself serves as the source of GWs. The peak frequency of the emitted GWs typically scales inversely with the mass $M_{\rm PBH}$ of the PBHs; for the low mass regime relevant for the PBHs that disappeared, and for certain formation mechanisms,  the frequency lies in the MHz-GHz range\footnote{While correlations between Hawking evaporation and GWs generated during the  formation of PBHs have been used to study stable PBHs~\cite{Agashe:2022jgk, Marfatia:2021hcp, Xie:2023cwi}, the method also holds great potential for lighter PBHs that evaporated away.  For instance, baryogenesis arising from the Hawking radiation of light PBHs has been studied by the current authors~\cite{Gehrman:2022imk}. The associated MHz-GHz GWs induced by curvature perturbations have been highlighted as a method to investigate the cosmology of PBHs, even if they have completely evaporated.}.

The purpose of this paper is to explore the signatures of PBHs in the MHz-GHz GW frontier, juxtaposed with their connection to DM. The overall scheme is as follows. PBHs are assumed to Hawking evaporate into a DM particle $\chi$ that couples to the Standard Model exclusively through gravity and yields the observed relic density. We remain agnostic to the specific nature of $\chi$. The correct relic abundance is obtained by a combination of three parameters: the mass and abundance of the PBHs ($M_{\rm PBH}$ and $\beta_{\rm PBH}$, respectively) and the mass of the DM particle ($m_\chi$). Given a  benchmark value of $m_\chi$, one then has a region in the plane $\{M_{\rm PBH}, \beta_{\rm PBH}\}$ where the correct relic density is achieved; in this same regime, one can calculate the correlated ultra-high frequency GWs. Ultimately, one has a DM-compatible map on the strain-frequency plane ($\{h_c, f_{\rm GW}\}$) in the ultra-high frequency GW frontier. 

The map from the plane characterizing the PBH properties $\{M_{\rm PBH}, \beta_{\rm PBH}\}$ to the plane characterizing GWs $\{h_c, f_{\rm GW}\}$ depends on the PBH formation mechanism. We explore two such formation mechanisms. The first is the canonical formation of PBHs by curvature perturbations during inflation~\cite{Carr:1975qj,Ivanov:1994pa,Garcia-Bellido:1996mdl,Silk:1986vc,Kawasaki:1997ju,Yokoyama:1995ex, Choudhury:2011jt,Choudhury:2013woa, Pi:2017gih,Hertzberg:2017dkh, Ozsoy:2018flq, Cicoli:2018asa}. Over-dense regions can collapse into a PBH when they enter the causal horizon. In this case, GWs originate from second-order effects. The same scalar perturbation responsible for the PBH formation contributes to tensor modes at horizon reentry. The map from the space of PBH properties to GWs is the following: $\{\Omega_\chi h^2,m_\chi\} \rightarrow \{M_{\rm PBH}, \beta_{\rm PBH}\} \, \rightarrow \{A_{\zeta}, k_p \} \rightarrow \{h_c, f_{\rm GW}\}$, where $A_\zeta$ is the amplitude of the power spectrum of the curvature perturbation and $k_p$ is the peak location of the power spectrum. The second formation mechanism we explore is the formation of PBHs from first-order phase transitions (FOPTs). The possibility of forming PBHs from collision of bubble walls during FOPTs has been studied for several decades \cite{Hawking:1982ga,Crawford:1982yz,Kodama:1982sf,Moss:1994pi,Freivogel:2007fx,Johnson:2011wt}. Here, we choose the more recently proposed mechanism of PBH formation from the collapse of particles trapped in the false vacuum \cite{Baker:2021nyl, Baker:2021sno, Kawana:2021tde,Jung:2021mku, Huang:2022him, Lu:2022paj}. The source of GWs in this case is the FOPT itself. The map from the space of PBH properties to GWs in this case is the following: $\{\Omega_\chi h^2,m_\chi\} \rightarrow \{M_{\rm PBH}, \beta_{\rm PBH}\} \, \rightarrow \{\alpha, \beta, T_\star, v_w \} \rightarrow \{h_c, f_{\rm GW}\}$, where $\alpha, \beta, T_\star$ and $v_w$ are the energy density released during phase transition normalized by the radiation energy density;  the inverse time scale of the phase transition; the temperature of the phase transition; and the bubble wall velocity, respectively.

One can further enquire: do the maps described above enable one to \textit{distinguish} between PBH formation mechanisms? In other words, given $\{\Omega_\chi h^2,m_\chi\}$, does one map to different points in the space of $\{h_c, f_{\rm GW}\}$ depending on the intermediate steps? The answer turns out to be in the affirmative, holding out the promise not only that ultra-high frequency GWs  will provide a connection between PBHs and DM, but also indicate the origin mechanism.  A fair criticism of the kind of precision study we are advocating is that the experimental status of ultra-high frequency GWs is not mature enough to be amenable to such studies yet. Our response is that this is a frontier of critical importance, as evidenced by the many ideas for probing it that have flowered recently \cite{Harry:1996gh, Arvanitaki:2012cn, Aggarwal:2020umq, Goryachev:2014yra, Page:2020zbr,Goryachev:2021zzn, Chou_2017, Akutsu:2008qv, Berlin:2021txa, Berlin:2022hfx, Herman:2022fau, Herman:2020wao, Domcke:2022rgu, Berlin:2023grv, Ito:2019wcb, Domcke:2020yzq}\footnote{Moreover, as we will see, CMB-Stage 4  experiments~\cite{CMB-S4:2016ple} will come close to probing at least some formation mechanisms.}.

This paper is structured as follows. In Section~\ref{sec:DM}, we review DM production via Hawking evaporation of PBHs. In Section~\ref{sec:Formation}, after describing two formation mechanisms of PBHs (from FOPTs and curvature perturbations), we calculate the GWs correlated with each formation mechanism. Our results, including the connections between PBH formation mechanism, DM production, and high frequency GWs, and the future prospects of detecting them, are discussed in Section~\ref{sec:HFGW}.

\section{Dark Matter from Hawking Radiation}
\label{sec:DM}

In this section, we compute the mass and abundance of light PBHs that Hawking evaporate to give the observed relic density of DM. We will be largely agnostic about the nature of the DM particle $\chi$, which could be fermionic or bosonic. We also discuss how the two different origin mechanisms of the PBHs  affect the required mass and abundance.  

\subsection{Particle Production through Hawking Radiation}
We begin with a discussion of the Hawking evaporation of PBHs into $\chi$. PBHs which originated from a radiation-dominated era acquire negligible spin due to the pressure of the radiation~\cite{DeLuca:2019buf}. Therefore,  we assume that all the PBHs are Schwarzschild (non-rotating).
As soon as PBHs form, they start to lose their mass through Hawking evaporation~\cite{Hawking:1975vcx}. Hawking radiation of a PBH of mass $M_\text{PBH}$ consists of all the particles in the spectrum that are lighter than the instantaneous horizon temperature of the PBH given by:
\begin{equation}
T_\text{PBH}(t)=\frac{M_\text{Pl}^2}{8\pi M_\text{PBH}(t)}.
\label{eq:temp}
\end{equation}
Ignoring the deviation of the Hawking radiation from the black body spectrum which is expressed as greybody factors~\cite{Page:1976df}, the energy spectrum
of the $\chi$ particle of mass $m_\chi$ with $g_\chi$ degrees of freedom is given by:
\begin{equation}
\frac{d^2u_\chi(E,t)}{dtdE}=\frac{g_\chi}{8\pi^2}\frac{E^3}{e^{E/T_\text{PBH}(t)}\pm1},
\label{eq:rate}
\end{equation}
($+$ for fermion emission and $-$ for boson emission) where $u_\chi(E,t)$ is the total radiated energy per unit area of the BH, and $E$ is the energy of the emitted particle.

Due to Hawking evaporation, the mass of a PBH formed at $t=t_i$ with the initial value of $M_\text{PBH}(t_i)$ evolves with time as:
\begin{equation}
M_\text{PBH}(t)=M_\text{PBH}(t_i)\left(1-\frac{t-t_i}{\tau_\text{PBH}}\right)^{1/3},
\end{equation}
where
\begin{equation}
\tau_\text{PBH}=\frac{10240\pi}{g_\star(T_\text{PBH})}\frac{M_\text{PBH}^3(t_i)}{M_\text{Pl}^4},
\end{equation}
is the lifetime of the PBH and $g_\star(T)$  counts the relativistic degrees of freedom at temperature $T$.

The rate of emission of the $\chi$ particle per energy interval can be expressed as
\begin{equation}
\frac{d^2N_\chi}{dtdE}=\frac{4\pi r_\text{S}^2}{E}\frac{d^2u_\chi}{dtdE},
\label{eq:numrate}
\end{equation}
where $ r_\text{S}=2M_\text{PBH}/M_\text{Pl}^2$ is the Schwarzschild radius 
of the PBH.
The total number of $\chi$ particles, provided that it is a boson ($B$), emitted over the PBH lifetime is obtained by integrating Eq.~\eqref{eq:numrate} over energy and time:
\begin{eqnarray}
N_\chi&=&\frac{120\,\zeta(3)}{\pi^3}\frac{g_\chi}{g_\star(T_\text{PBH})}\frac{M_\text{PBH}^2(t_i)}{M_\text{Pl}^2},~~~~~~T_\text{PBH}(t_i)>m_\chi,
\label{eq:numberlight}\\
N_\chi&=&\frac{15\,\zeta(3)}{8\pi^5}\frac{g_\chi}{g_\star(T_\text{PBH})}\frac{M_\text{Pl}^2}{m_
\chi^2},~~~~~~~~~~~~~~~T_\text{PBH}(t_i)<m_
\chi.
\label{eq:numberheavy}
\end{eqnarray}
The total number of fermionic species ($F$) is $N_F=\frac{3}{4}\frac{g_F}{g_B}N_B$.

\subsection{PBH Evaporation in a Radiation-Dominated Era}
The amount of DM produced via Hawking evaporation of PBHs in a radiation-dominated era can be evaluated by using conservation of entropy. The DM yield today, at $t_0$, is given by:
\begin{equation}
    Y_\chi=\frac{n_\chi(t_0)}{s(t_0)}=\frac{n_\chi(t_\text{eva})}{s(t_\text{eva})}\simeq N_\chi\frac{n_\text{PBH}(t_{i})}{s(t_{i})},
    \label{eq:YRD1}
\end{equation}
where $n_\chi(t)$ and $n_\text{PBH}(t)$ are the number densities of DM particles $\chi$ and PBHs, respectively. We presume there is no number changing process in the DM sector after PBH evaporation. $N_\chi$ is the total number of particles $\chi$ emitted by one PBH (Eq.~\eqref{eq:numberlight} and Eq.~\eqref{eq:numberheavy}), and $s(t)$ is the entropy density given by
\begin{equation}
s(T)=\frac{2\pi^2}{45}g_{*,s}(T)T^3,~~~~~~g_{*,s}(T)=\sum_B g_B\left(\frac{T_B}{T}\right)^3+\frac{7}{8}\sum_F g_F\left(\frac{T_F}{T}\right)^3.
\label{eq:entropy}
\end{equation}
The entropy from PBH evaporation is negligible when the energy density of PBHs remains a small fraction of the Universe.  
Assuming a fraction $\beta_{\rm PBH}$ of the energy density of the Universe collapses
into PBHs at the formation time, the yield of DM is therefore
\begin{equation}
    Y_\chi=\beta_{\rm PBH} N_\chi\frac{1}{M_{\rm PBH}}\frac{\rho_\text{rad}(t_{i})}{s(t_{i})}=\frac{3}{4}\frac{g_\star(T_i)}{g_{\star,s}(T_i)}\beta_{\rm PBH} N_\chi \frac{T_{i}(M_{\rm PBH})}{M_{\rm PBH}},
    \label{eq:YRD2}
\end{equation}
where $\rho_\text{rad}(t_{i})$ and $T_{i}(M_{\rm PBH})$ are the energy density and the temperature of the Universe at the PBH formation time respectively. The mass of PBHs basically follows the horizon mass at the formation time, but the exact relationship between PBHs' mass and the temperature of the Universe at the formation time depends on the formation mechanism (See Section~\ref{sec:Formation} for two examples).

The relic abundance of DM is obtained as:
\begin{equation}
\Omega_\chi=\frac{\rho_{\chi(t_0)}}{\rho_c}=\frac{m_\chi Y_\chi }{\rho_c}s(t_0),
\label{eq:relicdef}
\end{equation}
where $\rho_c(t_0)=1.0537\times 10^{-5}\, h^2\,\,\rm{GeV}~{\rm cm}^{-3}$, $s(t_0)=2891.2\left({T_0}/{2.7255 {\rm K}}\right)^3 \rm{cm^{-3}}$, and $h=0.674$ is scaling factor for Hubble expansion rate~\cite{Planck:2018vyg}.

While the energy density of 
radiation dilutes as $\rho_{\rm rad}(t)\propto a^{-4}$ with the expansion of the Universe, the energy density of PBHs decreases as the energy density of non-relativistic matter, i.e., $\rho_{\rm PBH}(t)\propto a^{-3}$, where $a$ is the scale factor. 
Since $\rho_{\rm PBH}(t)/\rho_{\rm rad}(t)\sim a$, a population of PBHs formed within a radiation-dominated era may lead to a transition to an early matter-dominated epoch before they evaporate. If PBHs are abundant enough to cause an early matter-dominated epoch, then at some time, $t_{\rm early-eq}$, which has to be before PBH evaporation ($t_{\rm early-eq}\lesssim \tau_{\rm PBH}$), they come to dominate the energy density of the Universe: $\rho_{\rm PBH}(t_{\rm early-eq})/\rho_{\rm rad}(t_{\rm early-eq})\sim 1$. The critical initial abundance of PBHs, $\beta_c$ that can initiate an early matter-dominated epoch is related to the temperature of the Universe at the formation time, $T_i$:
\begin{equation}
\frac{\rho_\text{PBH}(T_\text{early-eq})}{\rho_\text{rad}(T_\text{early-eq})}=\frac{\rho_\text{PBH}(T_i)}{\rho_\text{rad}(T_i)}\frac{T_i}{T_\text{early-eq}}=\beta_c\frac{T_i}{T_\text{early-eq}}\sim 1.
\end{equation}
An early matter-dominated epoch caused by PBHs requires $\beta\gtrsim \beta_c$ with $\beta_c=T_{\rm eva}/T_i$ where $T_{\rm eva}$ denoted the temperature of the Universe at PBH evaporation time. If the energy density of PBHs ever dominated the energy density of the Universe, the reheating effect from PBH evaporation needs to be included when calculating the DM relic abundance. In this work, we focus on the scenario without early matter domination to avoid the dilution of GWs from PBH formation \footnote{GWs produced during a period of matter (PBH) domination has been considered in \cite{Inomata:2019ivs, Inomata:2020lmk, Papanikolaou:2020qtd,Papanikolaou:2022chm, Domenech:2020ssp, Domenech:2021wkk}.}.

\section{PBH Formation Mechanisms and Correlated Gravitational Waves}
\label{sec:Formation}

In this Section, we describe two formation mechanisms of PBHs: from first-order phase transitions, and from curvature perturbations \footnote{We note the PBH formation has also been studied in other mechanisms, such as scalar field fragmentations~\cite{Cotner:2016cvr,Cotner:2017tir,Cotner:2018vug,Cotner:2019ykd}, domain walls~\cite{Rubin:2000dq,Rubin:2001yw}, cosmic strings~\cite{Hawking:1987bn, Polnarev:1988dh,MacGibbon:1997pu, Brandenberger:2021zvn}, and metric preheating~\cite{Martin:2019nuw, Martin:2020fgl}. If GWs were generated in these mechanisms, one could also correlate light PBHs with the corresponding GW signals.}. In both cases, we provide calculations of the GWs correlated with the formation mechanism. 

\subsection{First-order Phase Transitions}

The formation of PBHs from FOPTs has a long history. The earliest mechanisms focused on PBH formation from the collision of bubble walls \cite{Hawking:1982ga,Crawford:1982yz,Kodama:1982sf,Moss:1994pi,Freivogel:2007fx,Johnson:2011wt}. Recent mechanisms have focused on the formation of PBHs by the collapse of matter or solitons in the false vacuum \cite{Baker:2021nyl, Baker:2021sno, Kawana:2021tde,Jung:2021mku, Huang:2022him, Lu:2022paj}. It is this avenue that we will explore.

A simple template for the particle physics sector responsible for the formation of PBHs as well as the phase transition consists of a scalar $\phi$ and a fermion $\psi$, interacting via a Yukawa coupling $y_{\psi}$:
\be
\mathcal{L} \, \supset \, \bar{\psi} \slashed{\partial}\psi \, - \, y_{\psi} \phi \bar{\psi} \psi  \, + \, \mathcal{L}_{SM,\chi}\,\,.
\ee
We will dub the sector containing $\{\phi, \psi\}$ as the phase transition or ``PT" sector. The scalar $\phi$ induces a phase transition with $\langle \phi \rangle=0$ in the false vacuum and $\langle \phi \rangle\gg T_{\star}$ in the true vacuum. We assume that the Standard Model sector and the $\{\phi, \psi\}$ sector are populated after reheating and evolve to the same temperature at the time of the phase transition: $T_
{\rm SM}=T_{\rm PT}=T_{\star}$. This can be achieved, for example, by coupling $\psi$ to the Standard Model through higher dimensional operators. We remain agnostic to the specific form of such couplings, since they do not influence the PBH formation process.

The mass of the fermion $\psi$ is determined by the vacuum expectation value of  $\phi$ as $m_{\psi}=y_{\psi} \langle \phi \rangle$. Since the kinetic energy of the fermions is $\mathcal{O}(T_{\star})$, requiring that the mass in the true vacuum $m_{\psi}\gg T_{\star}$ will ensure that most of the $\psi$ particles are trapped in the false vacuum. Indeed,  energy-momentum conservation ensures that the number of $\psi$ particles $n_{\psi}$ penetrating through the bubble wall with energy larger than $m_\psi$ is  Boltzmann suppressed: $n_{\psi} \propto e^{-m_{\psi}/T_{\star}}$.

Fermions trapped in the false vacuum subsequently collapse to form PBHs. However, their number density can be depleted by annihilation processes into the scalars $\phi$. A detailed treatment and delineation of the parameter space of masses, couplings, temperatures, bubble wall velocity  amenable to PBH formation was performed by the authors of \cite{Baker:2021nyl, Baker:2021sno}. Numerically solving the associated Boltzmann equations, it was determined that benchmark values of the coupling 
\be
y_\psi \, \sim \, 10^{-4} \sqrt{\frac{T_\star}{10^6 {\rm GeV}}}\,,
\ee
with bubble wall velocity $v_w = 0.5$ leads to successful PBH formation for a large range of phase transition temperatures. We assume that the trapping rate is $100\%$ in this study. In the following,  we denote both the fermion as well as the anti-fermion by $\psi$, except when we discuss their annihilation. 

Since our interest is in light PBHs, we will concentrate on high temperature phase transitions with $T_\star \, \sim \, \mathcal{O}(10^{11-16})$ GeV. Take $T_\star=10^{15}~{\rm GeV}$ as an example, the benchmark value of the coupling in such case turns out to be $y_\psi \, \sim \, 0.3$, with $m_\psi \sim 10\,T_\star$~\cite{ Baker:2021sno}. Spherical over-dense regions with radius $R_\star$ at the onset of the phase transition typically shrink by a factor of $\sim 3$ before they become smaller than the Schwarzschild radius, leading to PBH formation (right panel, Fig. 4 of \cite{Baker:2021sno}). We discuss this process at length below, highlighting the main results of \cite{Baker:2021sno} and ~\cite{Xie:2023cwi}.  

 The energy density of the false vacuum is dominated by the energy density of the trapped particles $\psi$, whose distribution follows
\bea
\rho_{\psi}(t) \simeq \rho_{\psi}^{\rm eq} \, \left( \frac{R_{\star}}{R(t)} \right)^{4}\,\,,
\eea
with $R(t)$ being the size of the trapped region at time $t$ and $\rho^{\rm eq}_{\psi}$ the equilibrium distribution of $\psi$. The evolution of the energy density therefore depends on the ratio of the initial pocket size $R_{\star}$  and the pocket size $R(t)$ at time $t$, with the scaling being quartic because $\psi$ number density is enhanced by the suppressed volume while $\psi$ particles are simultaneously being heated up by the wall\footnote{The total energy density of the pocket also receives contributions from Standard Model particles and is given by
\bea
\rho_{\rm pocket}(t)=\rho_{\rm sm}\left(\frac{t_{\star}}{t}\right)^2 + \rho_{\psi}^{\rm eq} \, \left( \frac{R_{\star}}{R(t)} \right)^{4}.
\eea
However, contributions from trapped $\psi$ particles dominate soon after $T_\star$. We do not include the contribution from the false vacuum energy density since it depends on the latent heat of the specific FOPT. Also, the PBH formation in our case occurs after the percolation of false vacuum pockets when vacuum energy has already been released to the radiation energy in $71\%$ of the spatial regions.}. The mass of the PBH, $M_{\rm PBH}$ and the corresponding abundance $\beta_{\rm PBH}$ can be calculated using the methods developed in~\cite{Xie:2023cwi}.  

The mass $M_{\rm PBH}$  is determined by the total mass of the false vacuum pocket at the moment of  gravitational collapse $t_{\rm collapse}$. The corresponding radius at $t_{\rm collapse}$ can be determined by requiring that the false vacuum pocket size equals the Schwarzschild radius $r_{s}$ of a black hole whose mass equals to the total energy inside the pocket: 
\bea
R(t_{\rm collapse}) = r_{s}= 2 \, G \, M_{\rm pocket} \simeq \frac{8\pi G}{3} \, \frac{R_{\star}^4}{r_s} \, \rho^{\rm eq}_{\psi},
\label{eq:FOPTCollapseRadius}
\eea
where $G=1/M_{\rm Pl}^2$ is the gravitational constant. 
The equilibrium distribution  $\rho^{\rm eq}_{\psi}$ is evaluated at the phase transition temperature $T_\star$. The relation between the final and initial pocket size can be estimated as
\bea
\frac{r_s}{R_{\star}}\simeq\sqrt{\frac{7\,g_{\psi}}{8 \, g_{\star}}} \, R_{\star} \, H_{\star}.
\label{eq:FOPTCollapseRadiusRatio}
\eea
Here $g_{\star}=g_{\star,{\rm sm}}+g_{\star,\phi}+g_{\star,\psi}=106.75+1+4\times(7/8)$ is the total number of relativistic degrees of freedom when both the Standard Model sector and the PT sector are in equilibrium.

At this point, a few comments are in order about the effect of $\psi\bar{\psi}$ annihilations on the PBH formation rate. The effect of annihilation is twofold. Firstly, annihilation dilutes the energy density in the false vacuum and thus decreases the mass of the final PBH. Secondly, for large annihilation rates, the PBH formation rate is suppressed if the Schwarzschild radius decreases faster than the pocket radius and Eq.~\eqref{eq:FOPTCollapseRadius} is never satisfied. The parametric dependence of the annihilation rate on the pocket radius is $n_{\psi}^2\propto R(t)^{-6}$. If false vacuum regions survive well after the phase transition begins, {\it i.e.} $R(t)\ll R_{\star}$, a significant portion of the $\psi$ energy density leaks into the true vacuum bubbles in the form of the annihilation final states $\phi$. Therefore, PBH formation requires $R_{\star}/r_{s}\sim\mathcal{O}(1)$, implying $R_{\star}\gtrsim H^{-1}_\star$. For example, a benchmark value of $R_{\star}=1.5 \, H_{\star}^{-1}$ is found to allow successful PBH formation by Boltzmann equation simulations in \cite{Baker:2021nyl, Baker:2021sno}. In this study, we assume the minimal initial pocket radius for PBH formation is $R_{\star}^{\rm min}=1.5 \, H^{-1}_\star$, below which the energy density evolution with active $\psi\bar{\psi}$ annihilation needs further inspections.

The energy density inside the pocket right before PBH formation can be calculated for arbitrary $R_\star$ as \footnote{We assume the dilution of energy density from the annihilation process is negligible for our FOPT model parameters. This is confirmed in the numerical simulation in Fig.~2 of \cite{Baker:2021nyl} where the increase of the trapped $\psi$ energy density can scale approximately as $R(t)^{-4}$ given that the annihilation cross section is not too large.}
\bea
\rho_{\rm pocket}(t_{\rm collapse})= \frac{g_{\star,{\rm sm}}+g_{\star,\phi}+g_{\star,\psi}\left(\sqrt{\frac{8g_{\star}}{7g_{\psi}}}\frac{H^{-1}_{\star}}{R_{\star}}\right)^4}{g_{\star,{\rm sm}}+g_{\star,\phi}+g_{\star,\psi}} \, \rho_\star \,\,.
\eea
Then, the mass of PBH from this pocket is 
\bea
M_{\rm PBH}&=&\frac{4\pi}{3} \, r_{s}^3 \, \rho_{\rm pocket}\nonumber\\
&=&\frac{m_{\rm pl}^2}{2}\left(\frac{7g_{\psi}}{8g_{\star}}\right)^{\frac{3}{2}} \left[\frac{g_{\star,{\rm sm}}+g_{\star,\phi}+g_{\star,\psi}\left(\sqrt{\frac{8g_{\star}}{7g_{\psi}}}\frac{H^{-1}_{\star}}{R_{\star}}\right)^4}{g_{\star,{\rm sm}}+g_{\star,\phi}+g_{\star,\psi}}\right] \left(\frac{R_\star}{H_{\star}^{-1}}\right)^6 H_{\star}^{-1}\nonumber\\
&\simeq& 0.5 \times \left(\frac{10^{15}~{\rm GeV}}{T_{\star}}\right)^{2} \left[1 + 32.8 \left( \frac{R_{\star}}{H_{\star}^{-1}} \right)^{-4}\right] \left(\frac{R_{\star}}{H_{\star}^{-1}}\right)^{6} {\rm g} \nonumber\\
&\simeq& 16.3 \times \left(\frac{10^{15}~{\rm GeV}}{T_{\star}}\right)^{2} \left(\frac{R_{\star}}{H_{\star}^{-1}}\right)^{2} {\rm g}.
\label{eq:MPBHFOPT}
\eea
In the last step, we assumed that the dominant energy density contribution is from trapped particles such that $\rho_{\rm pocket}\simeq\rho_{\psi}$. 
From Eq.~\eqref{eq:MPBHFOPT}, it is clear that the PBH mass is only dependent on $T_{\star}$ when $R_{\star}/H_{\star}^{-1}$ is fixed. Lighter PBHs are therefore produced when the phase transition  happens at a higher temperature. 

We can also write the PBH mass in terms of the horizon mass $M_H$,
\bea
M_{\rm PBH} &=& \left( \frac{g_{\star,\psi}}{g_{}\star}\right)^{\frac{3}{2}} \left[\frac{g_{\star,{\rm sm}}+g_{\star,\phi}+g_{\star,\psi}\left(\sqrt{\frac{8g_{\star}}{7g_{\psi}}}\frac{H^{-1}_{\star}}{R_{\star}}\right)^4}{g_{\star,{\rm sm}}+g_{\star,\phi}+g_{\star,\psi}}\right] 
 \left(\frac{R_\star}{H_{\star}^{-1}}\right)^6 M_{H} \nonumber\\
 &\simeq& 5.4\times 10^{-3}\times \left(32.8\left(\frac{R_{\star}}{{H^{-1}_{\star}}}\right)^{2}+\left(\frac{R_{\star}}{{H^{-1}_{\star}}}\right)^{6}\right) M_{H}.
\eea
For PBH formation during FOPT, the mass function peaks at $R_\star / H^{-1}_{\star} =1.5$, implying the mass ratio $\gamma\equiv\frac{M_{\rm PBH}}{M_H}$ has a typical value
\bea
\gamma_{\rm FOPT}\Big{|}_{ R_\star / H^{-1}_{\star} =1.5} \simeq 0.46.
\eea
The $\gamma_{\rm FOPT}$ value could be larger when the initial pocket radius is even larger than the horizon size~($\gamma_{\rm FOPT}\simeq 1$ when $R_{\star}/H_{\star}^{-1}=2$), but the probability of having a large false vacuum pocket is highly suppressed because of the persistent nucleation of new true vacuum bubbles.

The total number of PBHs that are formed from vacuum pockets is determined by the distribution of false vacuum regions whose initial radius satisfies $R_{\star}>R_{\star}^{\rm min}=1.5 \, H^{-1}_\star$. The PBH mass is determined by the initial radius of the pocket that later collapses into the PBH, and $M_{\rm PBH}$ increases with $R_\star$ in Eq.~\eqref{eq:MPBHFOPT}. The number density of remnant pockets at the false vacuum percolation time is found in \cite{Lu:2022paj} using the reverse time description,
\bea
\frac{dn_{\rm pocket}}{dR_{\star}} \simeq \frac{I_{\star}^4 \, \beta^4}{192 \, v_w^3}\, e^{4\beta R_{\star} / v_w-I_{\star} \, e^{\beta R_{\star}/v_w}} \left( 1 - e^{-I_{\star}e^{\beta R_{\star}/v_w}} \right).
\label{eq:FalseVacuumDistribution}
\eea
where $I_\star=-\ln{(0.29)}=1.238$. The parameters that enter the double exponential suppression are $R_\star$, $\beta$ and $v_w$. The suppression for large values of $R_\star$ implies that PBH formation is most efficient at the smallest allowed mass and drops quickly for heavier PBHs. Similarly, the suppression for large values of $\beta$ can be understood as follows: since a large nucleation rate would cause new broken-phase bubbles to appear in unbroken-phase regions, an original pocket would be broken up into separate small pockets in such a scenario, and the remaining smaller pockets fail to form PBHs because of the $\psi\bar{\psi}$ annihilation. This condition of a pocket radius successfully shrinking from $R_\star$ to the Schwarzschild radius without any additional true vacuum bubble seeded inside it gives the dominant suppression factor $e^{-I_\star \, e^{\beta R_\star/v_w}}$. PBH formation thus prefers small values of $\beta$ and large values of $v_w$. A small $\beta$ value can be realized with supercooling \cite{DelleRose:2019pgi}, where the order parameter of supercooled FOPTs can be larger $\langle\phi\rangle \gg T_\star$ in order to trap $\psi$ particles with a moderate Yukawa coupling strength.

\begin{figure}[h]
  \centering
    \includegraphics[width=0.46\textwidth]{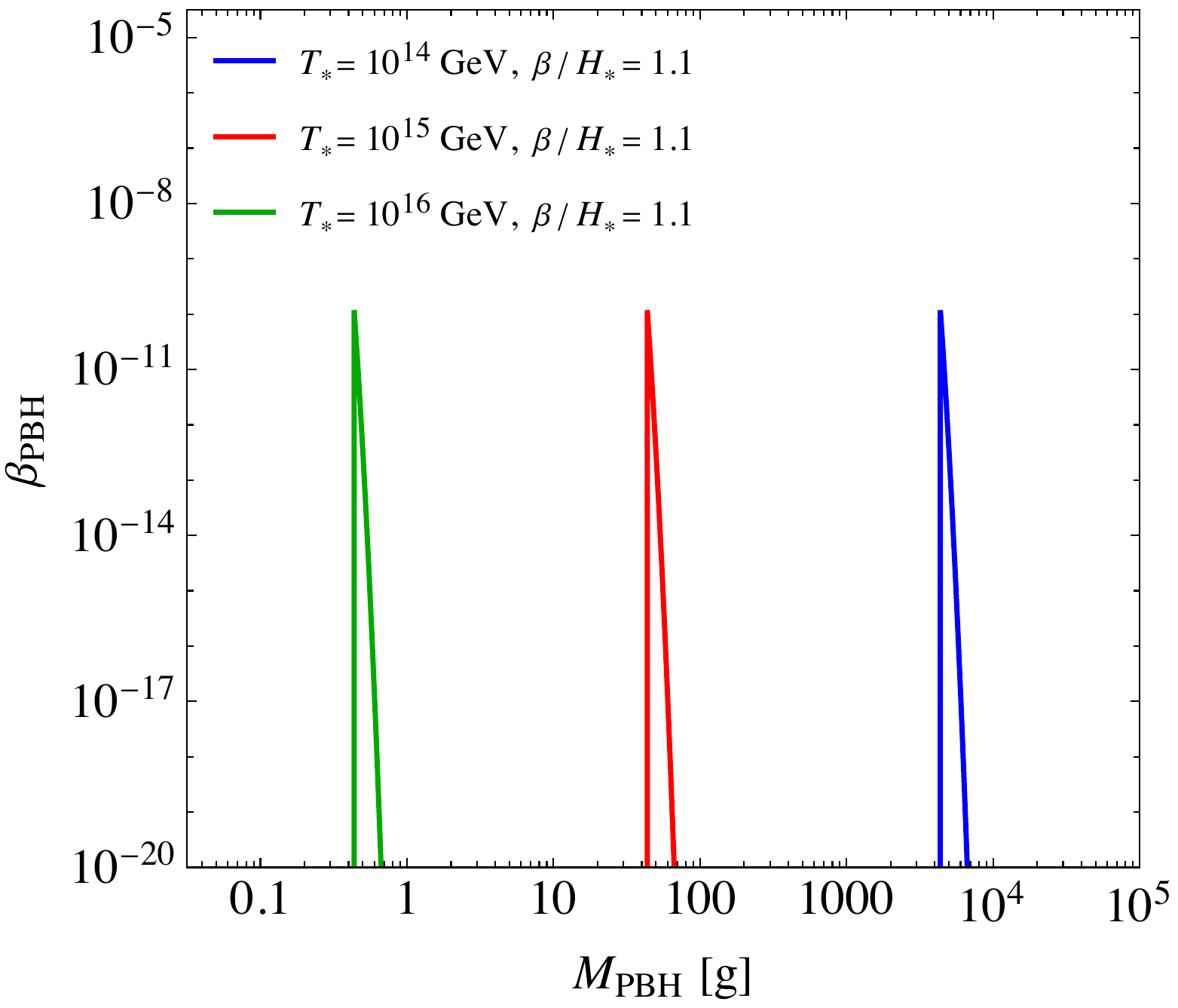} \qquad
    \includegraphics[width=0.47\textwidth]{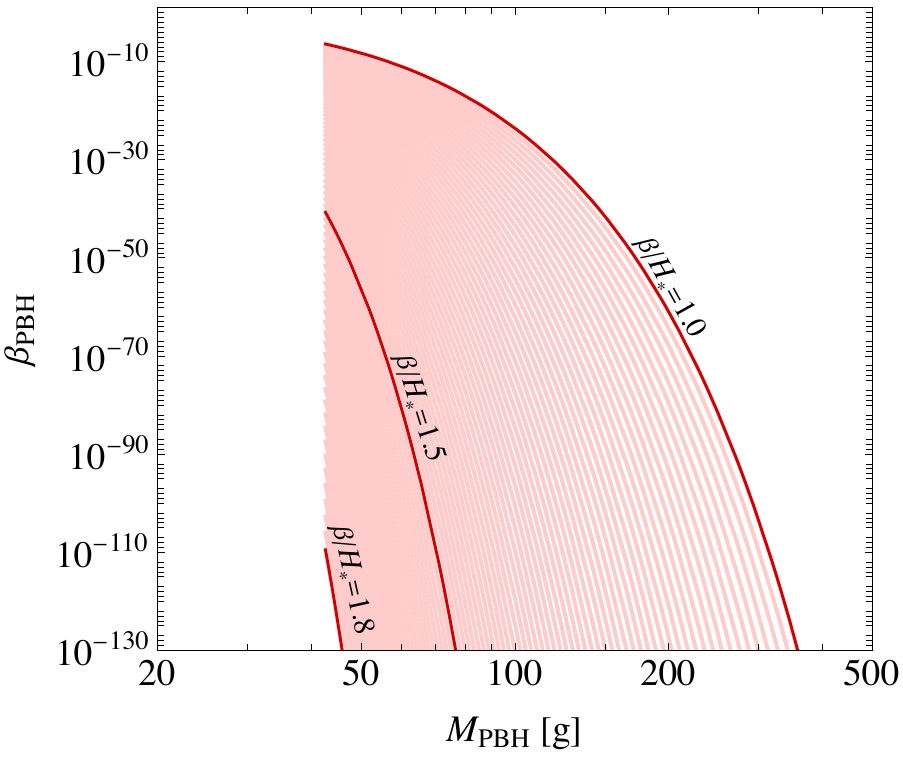} \\ 
    \includegraphics[width=0.46\textwidth]{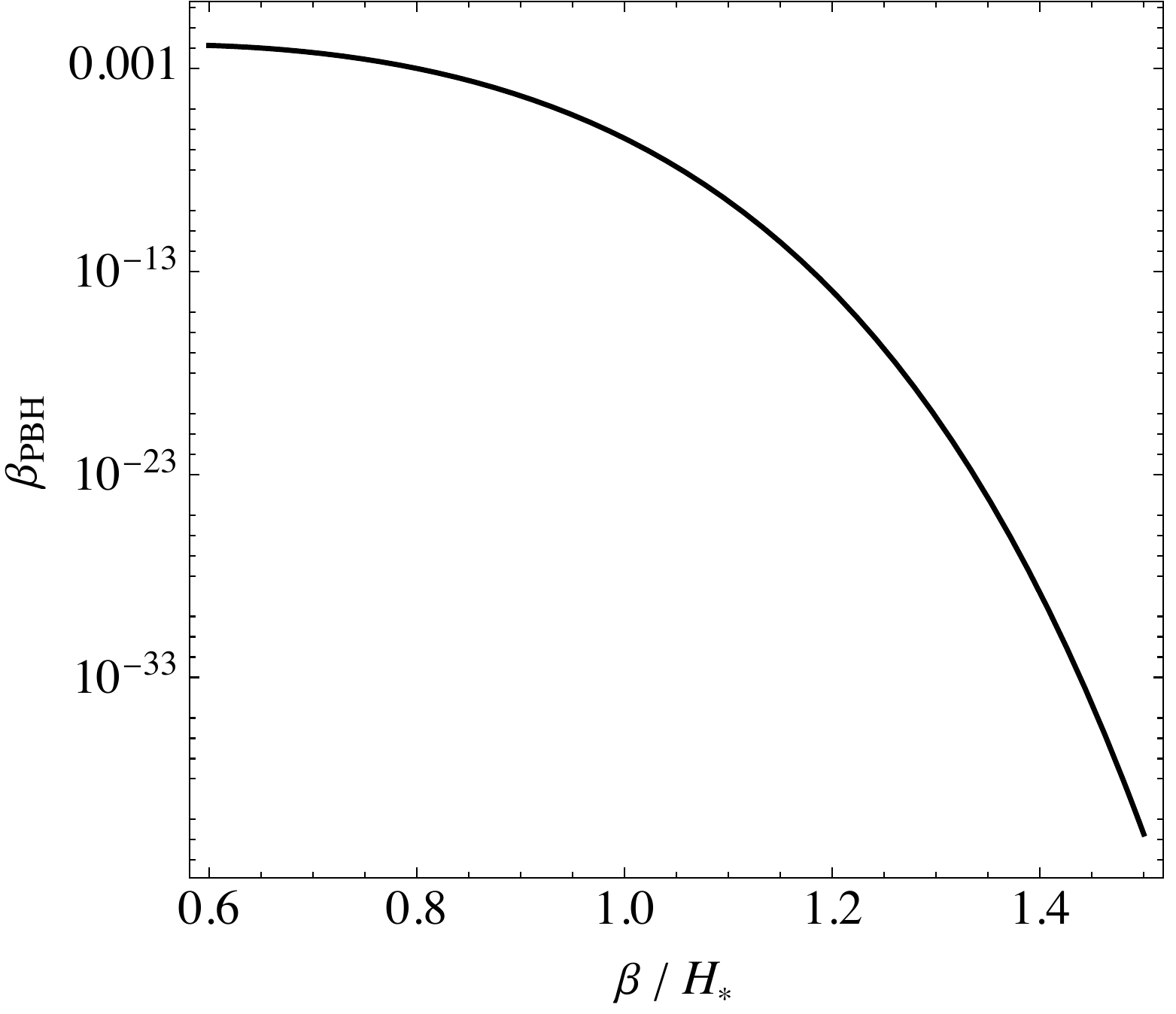}
  \caption{The PBH mass function generated by the FOPT formation mechanism. {\it Upper left panel:} The dependence on $T_\star$. FOPT temperatures are chosen $T_\star=10^{14}~{\rm GeV}$ (blue), $10^{15}~{\rm GeV}$ (red) and $10^{16}~{\rm GeV}$ (green), while $\beta/H_{\star}=1.1$ is fixed. The peak location of $\beta_{\rm PBH}$ is determined by $T_\star$ and the abundance is relatively unchanged. {\it Upper right panel:} The dependence on $\beta/H_{\star}$. The FOPT temperature is $T_\star=10^{15}~{\rm GeV}$. The range of $\beta_{\rm PBH}$ around the peak is shown for $\beta/H_{\star}\in[1.0,1.8]$. The PBH formation is suppressed with a larger $\beta$. {\it Lower panel:} The PBH abundance at peak location $R_\star/H_{\star}^{-1}=1.5$ with varying $\beta/H_\star$ values. The PBH formation rate is very sensitive to the increase of the nucleation rate.}\label{foptparamsbeta}
  \end{figure}

Each remnant pocket will collapse into a PBH, so that the PBH number density at  formation time can be written as
\bea
\frac{dn_{\rm PBH}}{dM_{\rm PBH}}&=&\frac{dR_{\star}}{dM_{\rm PBH}}\, \frac{dn_{\rm pocket}}{dR_{\star}}\Big{|}_{R_{\star}> 1.5 H_{\star}^{-1}}\nonumber\\
&=&\frac{R_\star}{2 \, M_{\rm PBH}} \, \frac{dn_{\rm pocket}}{dR_{\star}}\Big{|}_{R_{\star}> 1.5 H_{\star}^{-1}}.
\label{eq:dnPBHdMPBH}
\eea
The corresponding energy density fraction of PBHs is 
\bea
\beta_{\rm PBH}(M_{\rm PBH})=M_{\rm PBH} \frac{d\beta_{\rm PBH}}{dM_{\rm PBH}}=\frac{M_{\rm PBH}^2}{\rho_{\rm rad}(T_{\star})}\frac{dn_{\rm PBH}}{dM_{\rm PBH}}.
\label{eq:betaPBHFOPT}
\eea
Using Eq. \eqref{eq:FalseVacuumDistribution}-\eqref{eq:betaPBHFOPT}, and the $R_\star$-$M_{\rm PBH}$ relation in Eq. \eqref{eq:MPBHFOPT}, one can get
\bea
\beta_{\rm PBH}=\frac{M_{\rm PBH} R_{\star}}{2 \rho_{\rm rad}}\frac{I_{\star}^4 \, \beta^4}{192 \, v_w^3}\, e^{4\beta R_{\star} / v_w - I_{\star} \, e^{\beta R_{\star}/v_w}} \left( 1 - e^{-I_{\star}e^{\beta R_{\star}/v_w}} \right) \Big{|}_{R_{\star}> 1.5 H_{\star}^{-1}}. \label{eq:betapbhfinalfull}
\eea

At this point, we are in a position to obtain the PBH mass spectrum ($\beta_{\rm PBH}$ vs. $M_{\rm PBH}$ for fixed values of the FOPT parameters) as well as points on the  plane of $\{M_{\rm PBH}, \beta_{\rm PBH}\}$  as different selections of FOPT parameters  $\{\beta, T_\star, v_w \}$ are varied, following Eq.~\eqref{eq:MPBHFOPT} and Eq.~\eqref{eq:betapbhfinalfull}. The results are displayed in Fig.~\ref{foptparamsbeta}. On the upper left panel of Fig.~\ref{foptparamsbeta}, we show the PBH mass spectrum for fixed values of $T_\star = 10^{14}$ GeV (blue), $T_\star = 10^{15}$ GeV (red), and $T_\star = 10^{16}$ GeV (green). The wall velocity is fixed at $v_w = 0.5$, while $\beta$ is fixed at $\beta/H_\star = 1.1$. The PBH mass function generated through trapping particles during a FOPT is featured with a very sharp peak. The smaller PBH mass region is cut by the requirement $R_\star^{\rm min}=1.5 H^{-1}_{\star}$ to avoid excessive annihilation. The mass function for larger PBH masses is suppressed by the number density of larger pockets in Eq. \eqref{eq:FalseVacuumDistribution}. On the upper right panel of Fig.~\ref{foptparamsbeta}, we  fix value   $T_\star = 10^{15}$ GeV and $v_w = 0.5$, and vary $\beta/H_\star = 1.0 - 1.8$. In the lower panel, we show the amplitude of the PBH mass function peak as a function of the $\beta$ parameter. Larger $\beta$ means a higher nucleation rate and therefore the PBH mass function is suppressed.  

\begin{figure}[h]
  \centering
    \includegraphics[width=0.5\textwidth]{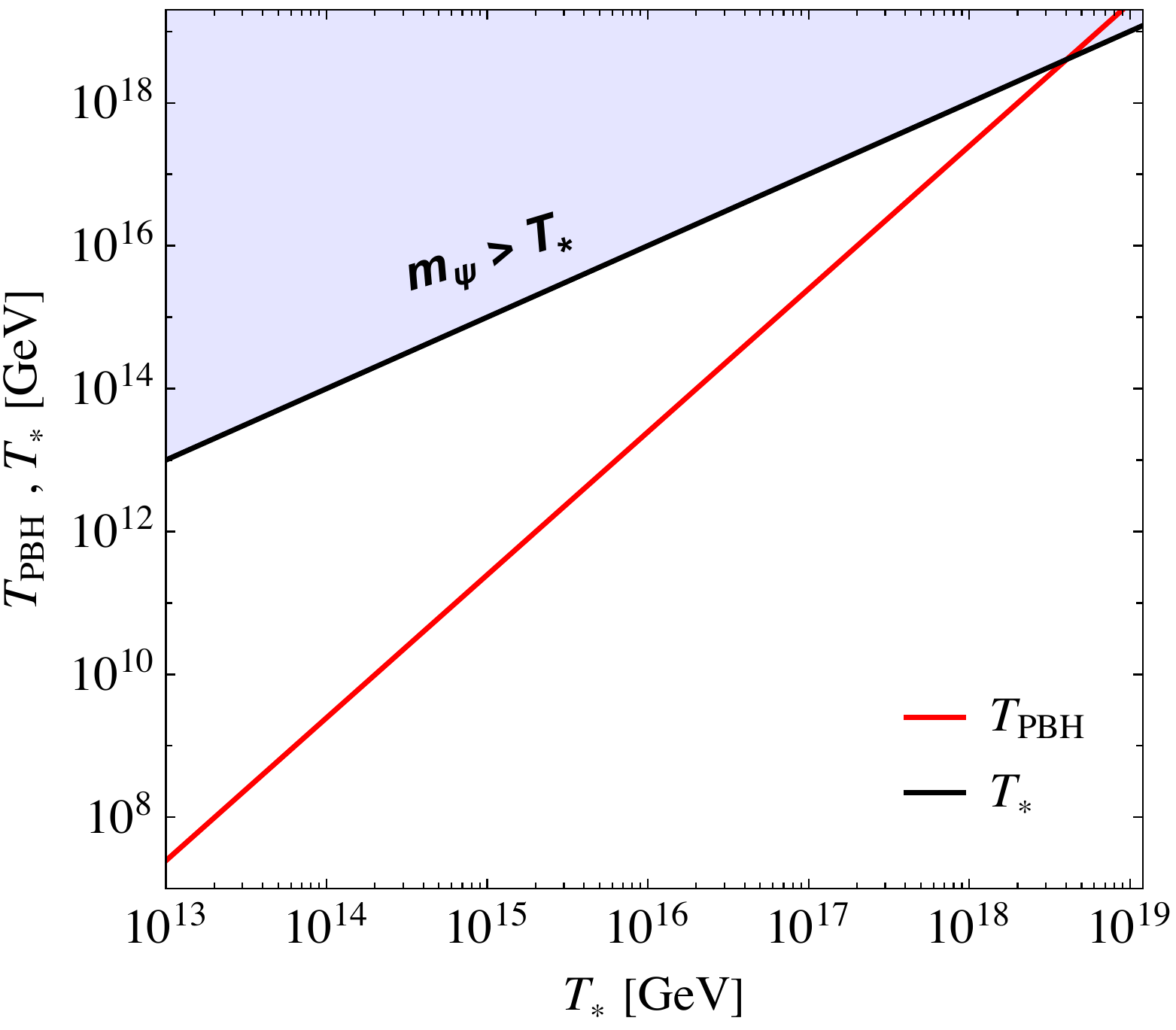}
  \caption{The relation between the FOPT temperature $T_{\star}$ (black) and the Hawking temperature $T_{\rm PBH}$ (red) of PBHs formed during a FOPT at $T_{\star}$, assuming the initial pocket radius is $R_{\star}=1.5 H^{-1}_{\star}$. The blue shaded region indicates that the $\psi$ particle mass in the true vacuum should at least be above the black curve such that it can be trapped inside the false vacuum pocket. For $T_\star\lesssim 10^{18} {\rm GeV}$, the PBH Hawking temperature is always smaller than $m_{\psi}$.}
  \label{fig:TPBHvsTstar}
\end{figure}

Before ending our discussion of PBH formation, we make a few comments about the subsequent Hawking evaporation of $\psi$. The Hawking production rate of $\psi$ is determined by the relation between $m_\psi$ and $T_{\rm PBH}$. In Fig.~\ref{fig:TPBHvsTstar}, we show $T_{\rm PBH}$ as a function of the formation time $T\simeq T_{\star}$. The requirement $m_\psi > T_\star$ is shown in blue. It is evident that for $T_\star \lesssim 10^{18} {\rm GeV}$, the PBH temperature is always smaller than the particle mass so that the Hawking radiation rate of $\psi$ is highly suppressed. Since $T_{\rm PBH}$ is also below the phase transition scale, $\psi$ is produced at the PBH horizon as a massive degree freedom in the true vacuum. We assume that massive $\psi$ particles that  penetrated into the true vacuum during the FOPT or were produced by  Hawking radiation of PBHs after the FOPT rapidly decay into the Standard Model, depleting any $\psi$ abundance.

We now turn to  a discussion of the GW signals correlated with PBH formation. We follow~\cite{Guo:2020grp,Guo:2021qcq} for the calculation of the GW signal from a FOPT, and focus on sound waves as the dominant contribution to the GW spectrum~\cite{Hindmarsh:2013xza, Hindmarsh:2015qta, Hindmarsh:2017gnf}. 
The peak frequency is given  by 
\bea
f^{\rm peak}_{\rm GW, sw}\simeq1.9\times10^{2} \, \left(\frac{1}{v_w}\right) \, \left(\frac{\beta}{H_{\star}}\right) \, \left(\frac{T_\star}{10^{15}~{\rm GeV}}\right) \, \left(\frac{g_\star(T_\star)}{106.75}\right)^{\frac{1}{6}}~{\rm MHz}.
\label{eq:fGWpeakFOPT}
\eea
The GWs from high temperature FOPTs that generate light PBHs have very high frequency. For $M_{\rm PBH}=0.1~{\rm g}$, the FOPT temperature is $T_\star\simeq 2\times 10^{16}~{\rm GeV}$, and the corresponding GW peak frequency is $f^{\rm peak}_{\rm GW, sw}\simeq 8.7~{\rm GHz}$ for the benchmark FOPT parameters discussed above. For the heaviest PBH mass we consider in this work, $T_\star\simeq 2\times 10^{11}~{\rm GeV}$ for $M_{\rm PBH}=10^9~{\rm g}$, the GW peak  frequency is $f^{\rm peak}_{\rm GW, sw}\simeq 0.09~{\rm MHz}$.
The FOPT GW energy density today is
\bea
\Omega_{\rm GW, sw} (f_{\rm GW}) &\simeq& 5.3 \times 10^{-6} \, \left(\frac{\beta}{H_{\star}}\right)^{-1} \, \left(\frac{\kappa \alpha}{1+\alpha} \right)^2 \, \left(\frac{106.75}{g_{\star}(T_\star)} \right)^{\frac{1}{3}} \, v_{w} \,  \Upsilon \nonumber\\ && \quad \times \left( \frac{f_{\rm GW}}{f^{\rm peak}_{\rm GW}} \right)^3 \, \left( \frac{7}{4 + 3 (f_{\rm GW}/f^{\rm peak}_{\rm GW)})} \right)^{\frac{7}{2}},
\label{eq:OmegaGWFOPT}
\eea
where $\Upsilon$, the suppression factor coming from the finite lifetime of sound waves, is given by~\cite{Guo:2020grp}:
\bea
\Upsilon=1-\frac{1}{\sqrt{1+2\tau_{\rm sw}H_{\star}}}.
\label{eq:FOPTGWUpsilon}
\eea
The lifetime of source can be calculated as
\bea
\tau_{\rm sw}\simeq \frac{R_\star}{\bar{U}_f},
\eea
where the $\bar{U}_f$ is the root mean square fluid velocity which is obtained as
\bea
\bar{U}^2_f=\frac{3}{4} \, \kappa \alpha.
\eea
The $\kappa$ parameter is the fraction of vacuum energy that is released during the phase transition which goes into the fluid motion \cite{Espinosa:2010hh}. The $\Omega_{\rm GW,sw}$ is inversely proportional to $\beta$, which means the GW signal is stronger for slow FOPTs that produced PBHs. Although we fix the bubble wall velocity to $v_w=0.5$, the GW signal is also enhanced by higher wall velocities as long as the $\psi$ particle trapping rate is not significantly affected.

\subsection{Scalar Perturbations}

In this Section, we discuss the second PBH formation mechanism relevant for us: primordial scalar perturbations. PBH formation from primordial scalar perturbations has been studied in~\cite{Carr:1975qj,Ivanov:1994pa,Garcia-Bellido:1996mdl,Silk:1986vc,Kawasaki:1997ju,Yokoyama:1995ex, Choudhury:2011jt,Choudhury:2013woa, Pi:2017gih,Hertzberg:2017dkh, Ozsoy:2018flq, Cicoli:2018asa}. These perturbations can be generated during the inflation era when a temporary ultra slow-roll phase~\cite{Tsamis:2003px, Kinney:2005vj, Martin:2012pe, Motohashi:2014ppa,Anguelova:2017djf, Dimopoulos:2017ged} enhances the curvature perturbation power spectrum.
In this work, we use a $\delta$-function shape for the power spectrum of the curvature perturbation for illustration of our idea,
\bea
P_{\zeta}(k)=A_\zeta \, \delta(\log k - \log k_p),
\label{eq:deltaPzeta}
\eea
where $k_p$ determines the scale of the enhanced perturbation and $A_\zeta$ determines the amplitude of the perturbation.

These primordial fluctuations are frozen at super-horizon scales after they are generated during inflation and enter the causal horizon at later time as a result of the cosmic expansion. When the over-density enters the horizon, the gravitational attraction could overcome the pressure and leads to the gravitational collapse of a proportion of the Hubble patch into a PBH. The black hole formation requires the local over-density surpasses a threshold value, i.e., $\delta>\delta_c$. The value of the threshold for a radiation-dominated Universe is $\delta_c\simeq\omega=1/3$ 
where
$\omega=p/\rho$ is the equation of state parameter~\cite{Carr:1975qj}. We assume the distribution of $\delta$ in all patches is Gaussian:
\bea
p(\delta)=\frac{1}{\sqrt{2\pi}\sigma_0} \, e^{-\frac{\delta^2}{2\sigma^2_0}}.
\label{eq:deltagaussian}
\eea
The mean value of $\delta$ is zero and the variance $\sigma^2$ is calculated from the curvature perturbation as
\bea
\sigma_0^2(k=R^{-1})=\displaystyle{\int_0^{\infty}}\frac{{\rm d}k'}{k'}\frac{16}{81}(k'R)^4W^2(k',R)P_\zeta(k').
\label{eq:sigma0}
\eea
If the curvature perturbation takes the monochromatic form in Eq.~\eqref{eq:deltaPzeta}, the variance is simplified to
\bea
\sigma^2_0(k)=A_\zeta \, \frac{16}{81} \, 
\left(\frac{k_p}{k}\right)^4 \, \exp\left[-\left(\frac{k_p}{k}\right)^2\right].
\label{eq:Sigma0sqDeltaPS}
\eea
The variance is generated at a range of length scales $R\sim k_{p}^{-1}$ even if the curvature perturbation itself is monochromatic at $k_p$. This is because fluctuations are coarse-grained averaged with the window function
\bea
W(k,R)=\exp\left[-\frac{(kR)^2}{2}\right].
\eea
The window function suppress contributions from perturbations at much smaller length scales. 
We define the abundance of PBHs in the same way as used in the FOPT case. Moreover, we assume the fraction of the energy density collapsed into the PBH is $\gamma=0.2$,
\bea
M_{\rm PBH}(k)=\gamma \, M_{H}(k) = 4 \times 10^{3} \left(\frac{\gamma}{0.2}\right) \left( \frac{k}{10^{21} {\rm Mpc}^{-1}} \right)^{-2} \, {\rm g}.
\label{eq:MPBHPzeta}
\eea
Note that the ratio of the PBH mass and the horizon mass at formation time in this mechanism, $\gamma=0.2$, is about half of the typical ratio value of the FOPT mechanism. The energy density of PBHs can be estimated with the PBH mass and the PBH number density. The number density is equal to the number density of horizon patches that have large enough density contrasts for them to undergo gravitational collapse, $n_{\rm PBH}=n_{\rm patch}|_{\delta>\delta_c}$, which can be calculated by integrating the Gaussian distribution in Eq. \eqref{eq:deltagaussian} from $\delta_c$ to infinity as
\bea
\beta_{\rm PBH}&=&\gamma \, \int_{\delta_c}^{\infty}d\delta \, \frac{1}{\sqrt{2\pi}\sigma_0}e^{-\frac{\delta^2}{2 \sigma_0^2}}\nonumber\\
&=&\frac{\gamma}{2} \, {\rm Erfc}\left(\frac{\delta_c}{\sqrt{2}\sigma_0}\right).
\label{eq:betaPBHErfc}
\eea
\begin{figure}[h]
  \centering
   \includegraphics[width = 0.46 \textwidth]{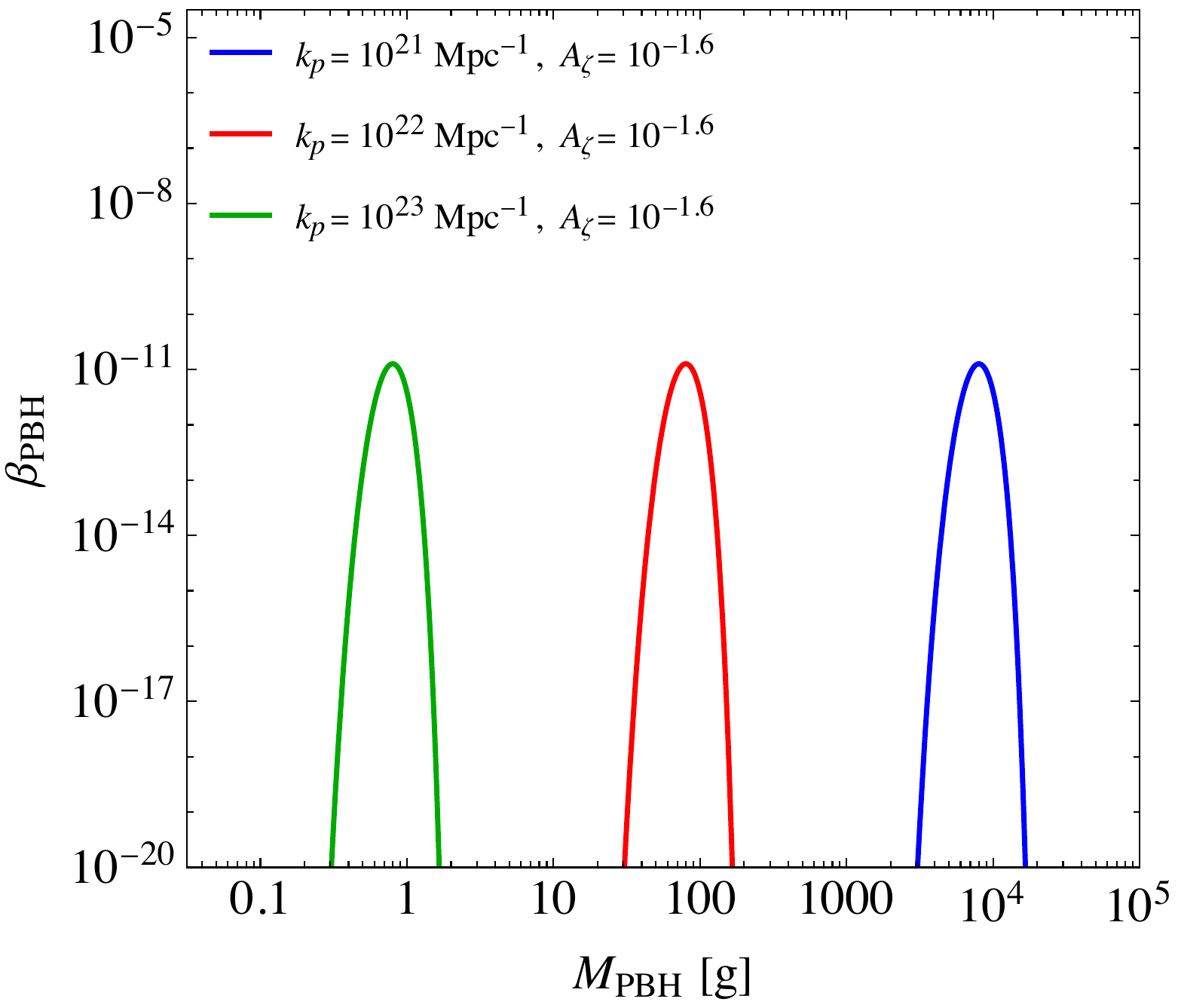}\quad 
    \includegraphics[width = 0.5 \textwidth]{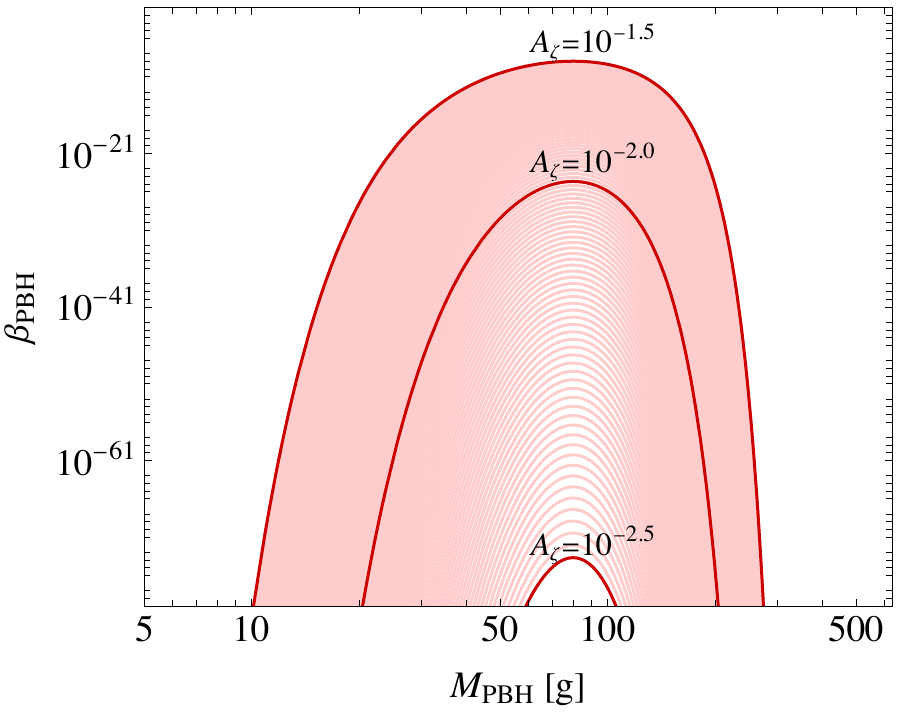}\\
     \includegraphics[width = 0.46 \textwidth]{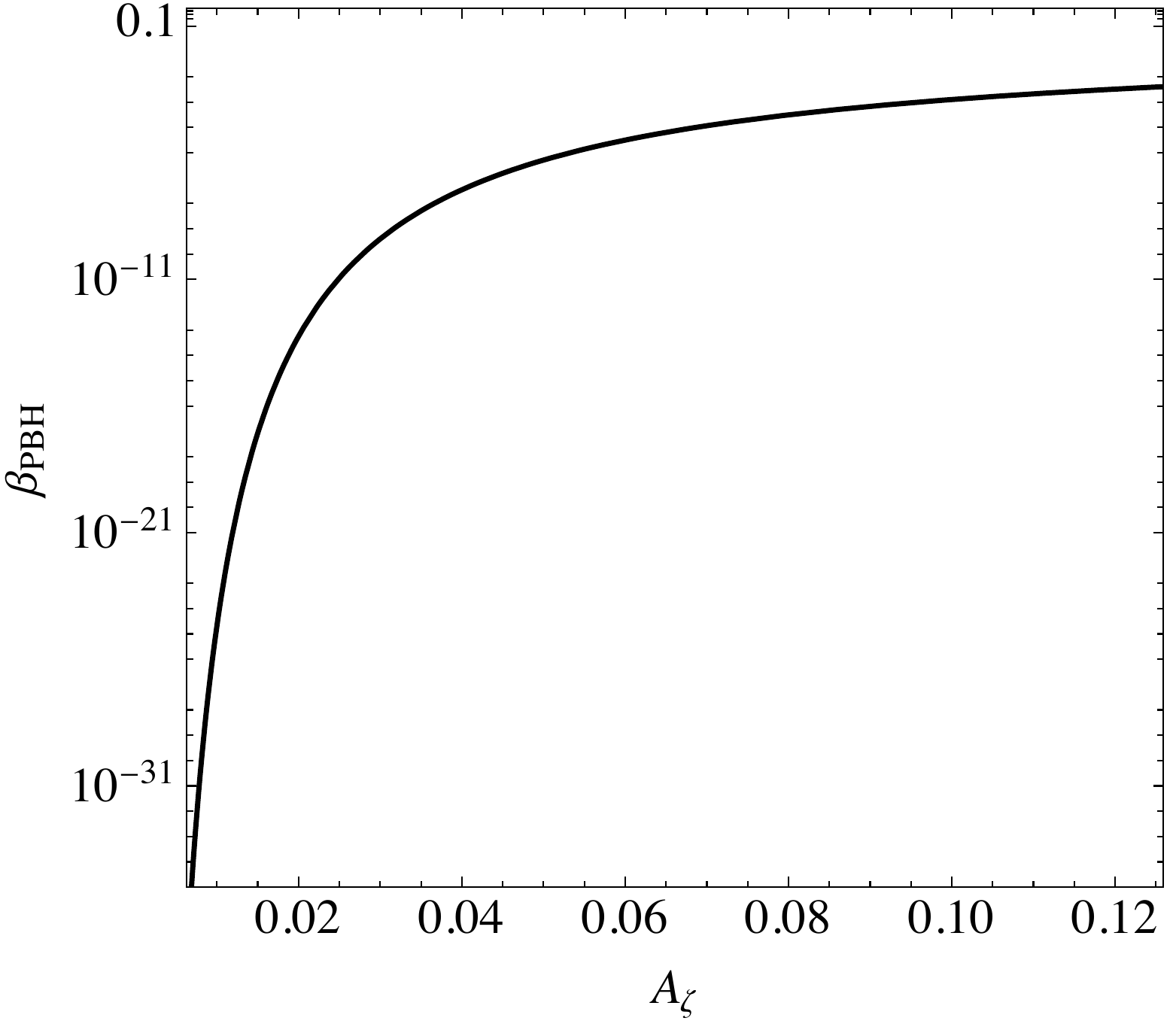}
     \caption{The PBH mass function generated by primordial perturbations. {\it Upper left panel:} The dependence on $k_p$. The scales of the fluctuation are chosen $k_p=10^{21}~{\rm Mpc}^{-1}$ (blue), $10^{22}~{\rm Mpc}^{-1}$ (red) and $10^{23}~{\rm Mpc}^{-1}$ (green), while the amplitude $A_\zeta=10^{-1.6}$ is fixed. The peak location of $\beta_{\rm PBH}$ is determined by $k_p$ and the abundance is unchanged. {\it Upper right panel:} The dependence on $A_\zeta$. The peak $k$-mode is $k_p=10^{22}~{\rm Mpc}^{-1}$. The range of $\beta_{\rm PBH}$ around the peak is shown for $A_\zeta\in[10^{-2.5},10^{-1.5}]$. {\it Lower panel:} The PBH abundance at peak location with varying $A_\zeta$ values. The PBH formation rate drops quickly when the variance of the density contrast distribution becomes smaller than the formation threshold.}
  \label{fig:BetaMassFrac}
\end{figure}
In Fig.~\ref{fig:BetaMassFrac}, we show example PBH mass functions $\{M_{\rm PBH}, \beta_{\rm PBH}\}$ generated by primordial scalar perturbations $\{ A_{\zeta}, k_p\}$. On the upper left panel of Fig.~\ref{fig:BetaMassFrac}, we show the PBH mass spectrum for fixed values of power spectrum peak locations $k_{p}=10^{21}~{\rm Mpc}^{-1}$ (blue), $k_{p}=10^{22}~{\rm Mpc}^{-1}$ (red) and $k_{p}=10^{23}~{\rm Mpc}^{-1}$ (green). The amplitude of the power spectrum is chosen to be $A_\zeta=10^{-1.6}$. The peak of the $M_{\rm PBH}$ distribution is determined by $k_p$ with Eq.~\eqref{eq:MPBHPzeta}. On the upper right panel of Fig.~\ref{fig:BetaMassFrac}, we show mass functions for a range of $A_\zeta$ values with fixed $k_p=10^{22}~{\rm Mpc}^{-1}$. In the lower panel, we show the peak $\beta_{\rm PBH}$ value as a function of $A_\zeta$. The variance of $\delta$ decreases with smaller $A_\zeta$ in Eq.\eqref{eq:Sigma0sqDeltaPS} such that the PBH formation rate is suppressed with small amplitudes of primordial fluctuations.

Besides the PBH formation, GWs are induced by the second-order effect when scalar modes enter the particle horizon at $a H=k_p$~\cite{Ananda:2006af,Baumann:2007zm,Assadullahi:2009jc,Acquaviva:2002ud,Kohri:2018awv,Inomata:2018epa,Domenech:2021ztg}. We follow~\cite{Inomata:2018epa, Kozaczuk:2021wcl, Agashe:2022jgk} for the calculation of such induced GWs. The GW density today $\Omega_{{\rm GW},0}$ is expressed as
\bea
\Omega_{{\rm GW},0}(f_{\rm GW})=0.39 \, \left( \frac{g_{\star}(\eta_s)}{106.75} \right)^{-\frac{1}{3}} \, \Omega_{{\rm rad},0} \, \Omega_{{\rm GW},\zeta}(\eta_s,k),
\eea
where the GW frequency and the $k$-mode is related by $f_{\rm GW}=1.546\times10^{-15}({k}/{{\rm Mpc}^{-1}})~{\rm Hz}$. The $\Omega_{{\rm rad},0}=8.5\times 10^{-5}$ is the energy density of radiation today normalized to the critical density. The GW density at a conformal time $\eta_s$ subsequent to the horizon reentry is~\cite{Inomata:2018epa}
\bea
\Omega_{{\rm GW},\zeta}(\eta,k)=\frac{1}{24}\left(\frac{k}{a(\eta) \, H(\eta)}\right)^2P_h(\eta,k),
\label{eq.GW}
\eea
where the power spectrum of tensor modes $P_h(\eta,k)$ is calculated with
\bea
P_h(\eta,k)&\simeq&2\int^{\infty}_{0}dt\int^{1}_{-1}ds\left(\frac{t(t+2)(s^2-1)}{(t+s+1)(t-s+1)}\right)^2\nonumber\\
&&\times I^2(s,t,k\eta)P_\zeta(u k)P_\zeta(v k).
\label{eq:GWPh}
\eea
The dimensionless variables are defined $u=\frac{t+s+1}{2}$ and $v=\frac{t-s+1}{2}$. In a radiation-dominated Universe, sub-horizon perturbation modes decay quickly due to pressure after horizon re-entry. Therefore GWs are mostly produced at the re-entry time and evolve to constant values in the sub-horizon limit. The $I^2$ term can be written within this limit as
\bea
I^2(s,t,k\eta)&=&\frac{288(s^2+t(t+2)-5)^2}{k^2\eta^2(t+s+1)^6(t-s+1)^6}\bigg(\frac{\pi^2}{4}\Big(s^2+t(t+2)-5\Big)^2\Theta(t-(\sqrt{3}-1))\\
&&\qquad\qquad+\Big(-(t+s+1)(t-s+1)+\frac{1}{2}(s^2+t(t+2)-5)\log\Big|\frac{t(t+2)-2}{3-s^2}\Big|\Big)^2\bigg)\nonumber,
\eea
where $\Theta(.)$ is the Heaviside function. After integrating out the two $\delta$-functions appeared in Eq.~\eqref{eq:GWPh} as a result of the second-order effect,
\bea
\Omega_{{\rm GW}, \zeta}=\frac{3}{64} A^2_\zeta \, &r^2& \, \left(\frac{4-r^2}{4}\right)^2 \, (2-3r^2)^2 \\
&\times&\left[\left(4+(3r^2-2)\log \left|\frac{3r^2-4}{3r^2}\right|\right)^2+\pi^2(3r^2-2)^2\Theta\left(\frac{2}{\sqrt{3}}-r\right)\right] \Theta(2-r)\nonumber,
\label{eq:GWzeta}
\eea
where $r=k/k_p$. For a general shape of the curvature perturbation power spectrum, the peak frequency of the induced GW spectrum is determined by the horizon reentry time of the power spectrum peak $k_p$ as:
\bea
f^{\rm peak}_{{\rm GW}, \zeta} = 1.546 \times \left(\frac{k_p}{10^{21} {\rm Mpc}^{-1}}\right) \, {\rm MHz}.
\label{eq:fGWpeakPzeta}
\eea
The induced GWs are generated at very high frequencies when large curvature perturbations occur at small scales. We calculate the GW peak frequency with Eq.~\eqref{eq:MPBHPzeta} and Eq.~\eqref{eq:fGWpeakPzeta}, $f^{\rm peak}_{{\rm GW}, \zeta}\simeq 0.3~{\rm GHz}$ for $M_{\rm PBH}=0.1~{\rm g}$ and $f^{\rm peak}_{{\rm GW}, \zeta}\simeq 3\times10^{-3}~{\rm MHz}$ for $M_{\rm PBH}=10^9~{\rm g}$. The GW energy density is found to be proportional to the square of the amplitude of the curvature perturbation at the formation time $\Omega_{\rm GW}(\eta_s)\simeq A_\zeta^2$~\cite{Inomata:2018epa}, thus one can estimate the present day GW signal strength to be $\Omega_{{\rm GW},0}\simeq A_\zeta^2 \times \Omega_{{\rm rad},0}\sim10^{-9}$.

\section{Results: Dark Matter and  High-frequency Gravitational Waves}
\label{sec:HFGW}

In this Section, we collect our results from PBH formation and calculate the correlated GWs. Our first step will be to discuss the  PBH  parameter space for a given relic density and DM mass (this step constitutes the map  $\{\Omega_\chi h^2,m_\chi\} \rightarrow \{M_{\rm PBH}, \beta_{\rm PBH}\}$ of our results). We will then compute the resulting GWs for  the two formation mechanisms (this step constitutes the maps $ \{M_{\rm PBH}, \beta_{\rm PBH}\} \, \rightarrow \{\alpha, \beta, T_\star, v_w \} \rightarrow \{h_c, f_{\rm GW}\}$  and $ \{M_{\rm PBH}, \beta_{\rm PBH}\} \, \rightarrow \{A_{\zeta}, k_p \} \rightarrow \{h_c, f_{\rm GW}\}$ of our results).

\subsection{Correlating Dark Matter Abundance: $\{\Omega_\chi h^2,m_\chi\} \rightarrow \{M_{\rm PBH}, \beta_{\rm PBH}\}$ }

We would like to understand the following question: given a mass of DM $m_\chi$ and a given value of the DM relic density  $\Omega_\chi h^2$, what is the corresponding regime of $\{M_{\rm PBH}, \beta_{\rm PBH}\}$ that is covered? We can use Eq.~\eqref{eq:YRD2} and Eq.~\eqref{eq:relicdef} for the yield and relic density. While the results for the two formation mechanisms are similar, we treat them separately. We calculate quantities for the two example formation mechanisms with conventions "FOPT" for phase transition and "$\zeta$" for curvatue perturbations.

In the case of FOPT, we have
\be
Y_{\chi, \rm FOPT} \,=\, \frac{3}{4}\beta_{\rm PBH}N_{\chi}\frac{T_{\star}}{M_{\rm PBH}}\,\,.
\ee
Solving for $T_\star$ in the third line of Eq.~\eqref{eq:MPBHFOPT}, the following expression can be attained
\be
     T_{\star} \simeq 7.1\times10^{14}\,\sqrt{32.8+\left(\frac{R_{\star}}{H_{\star}^{-1}}\right)^4}\left(\frac{R_{\star}}{H_{\star}^{-1}}\right)\left(\frac{1~{\rm g}}{M_{\rm PBH}}\right)^{\frac{1}{2}}~{\rm GeV}\,\,,
     \label{eq:Tstar}
\ee
To produce the relic density of a real scalar DM $\chi$, we use Eq.~\eqref{eq:Tstar} and the relic density is given as
\bea
    \Omega_{\chi}h^{2}\,\simeq\,2.5 &\times& 10^{10} \, g_{\chi}\left(\frac{106.75}{g_{\star}(T_{\rm PBH})}\right)\sqrt{32.8+\left(\frac{R_{\star}}{H_{\star}^{-1}}\right)^4}\left(\frac{R_{\star}}{H_{\star}^{-1}}\right) \nonumber \\
    &\times& \, \beta_{\rm {PBH}}\left(\frac{M_{\rm PBH}}{1~{\rm g}}\right)^{1/2} \left(\frac{m_{\chi}}{\rm {TeV}}\right), ~~~~~~~~ T_{\rm PBH} > m_{\chi},
\eea
and
\bea
    \Omega_{\chi}h^{2}\,\simeq\,2.7 &\times& 10^{18} \, g_{\chi}\left(\frac{106.75}{g_{\star}(T_{\rm PBH})}\right)\sqrt{32.8+\left(\frac{R_{\star}}{H_{\star}^{-1}}\right)^4}\left(\frac{R_{\star}}{H_{\star}^{-1}}\right) \nonumber \\
    &\times& \, \beta_{\rm{PBH}} \left(\frac{1~{\rm g}}{M_{\rm{PBH}}}\right)^{3/2}\left(\frac{10^{15}~{\rm GeV}}{m_\chi}\right), ~~~~~~~~ T_{\rm PBH}< m_{\chi}.
\eea
The critical abundance beyond which the Universe becomes PBH-dominated can be obtained as follows
\bea
\beta_{\rm{PBH,crit}}^{\rm{FOPT}}&\simeq& 1.7\times 10^{-5} \left[32.8+\left(\frac{R_\star}{H^{-1}_\star}\right)^4\right]^{-\frac{1}{2}} \left(\frac{R_\star}{H^{-1}_\star}\right)^{-1}\left(\frac{1~{\rm g}}{M_{\rm{PBH}}}\right)\left(\frac{g_{\star}(T_{\star})}{106.75}\right)^{\frac{1}{4}}\nonumber\\
&\simeq&1.9\times 10^{-6}\left(\frac{1~{\rm g}}{M_{\rm{PBH}}}\right)\left(\frac{g_{\star}(T_{\star})}{106.75}\right)^{\frac{1}{4}}\Big{|}_{ R_\star / H^{-1}_{\star} =1.5}.
\label{eq:betacFOPTRH}
\eea

For scalar perturbations, the yield of DM is given by:
\be
Y_{\chi, \zeta} \,=\, \frac{3}{4}\beta_{\rm PBH}N_{\chi}\frac{T_i(M_{\rm PBH})}{M_{\rm PBH}}\,\,,
\ee
where the temperature of the Universe at PBH formation time, $T_i(M_\text{PBH})$, is obtained from writing Eq.~\eqref{eq:MPBHPzeta} in terms of temperature as $M_\text{PBH}=(4\pi/3)\gamma\rho_\text{rad}(T_i)H^{-3}(T_i)$. Therefore we have:
\be
T_{i,\zeta}\simeq 4.3 \times 10^{15} \left(\frac{\gamma}{0.2}\right)^{\frac{1}{2}}\, \left(\frac{106.75}{g_\star(T_i)}\right)^{\frac{1}{4}} \, \left(\frac{1~{\rm g}}{M_{\rm PBH}}\right)^{\frac{1}{2}}~{\rm GeV},
\ee
is the temperature of the Universe when PBH formation is from the collapse of curvature perturbations.
The relic abundance of DM from the curvature perturbation formation mechanism are as follows:
\bea
\Omega_\chi h^2 \simeq 1.6 &\times& 10^{13}\left(\frac{g_*(T_i)}{106.75}\right)^{-1/4} \, \left(\frac{\gamma}{0.2}\right)^{1/2} \, \left(\frac{g_\chi}{g_*(T_\text{PBH})}\right)\nonumber\\
&\times& \beta_\text{PBH} \, \left(\frac{M_\text{PBH}}{\rm{1g}}\right)^{1/2} \, \left(\frac{m_\chi}{\rm{TeV}}\right), ~~~~~~~~ T_\text{PBH}>m_\chi,
\eea
and 
\bea
\Omega_\chi h^2 \simeq  1.8 &\times& 10^{21} \, \left(\frac{g_*(T_i)}{106.75}\right)^{-1/4} \, \left(\frac{\gamma}{0.2}\right)^{1/2} \, \left(\frac{g_\chi}{g_*(T_\text{PBH})}\right) \nonumber\\
&\times& \beta_\text{PBH} \, \left(\frac{\rm{1g}}{M_\text{PBH}}\right)^{3/2} \, \left(\frac{10^{15}\rm{GeV}}{m_\chi}\right), ~~~~~~~~T_\text{PBH}<m_\chi.
\label{eq:relic}
\eea
The corresponding critical PBH abundance beyond which the Universe becomes matter-dominated in the curvature perturbation mechanism is given by
\begin{equation}
\beta_{\rm {PBH, crit}}^{\zeta} \simeq 2.8\times10^{-6}\left(\frac{g_{\star}(T_{\rm BH})}{106.75}\right)^{1/2}\left(\frac{0.2}{\gamma}\right)\left(\frac{\rm{1g}}{M_{\rm PBH}}\right).
\label{eq:betacScalar}
\end{equation}

We summarize the required PBH abundance to generate the correct DM relic abundance as a function of the DM mass and the PBH mass, for the two formation mechanisms. For PBH formation during FOPT,
\bea
\nonumber\beta_{\rm PBH}&=&\frac{1}{g_{\chi}}\left(\frac{g_{\star}(T_{\rm PBH})}{106.75}\right)\left[164+5\left(\frac{R_{\star}}{H_{\star}^{-1}}\right)^4\right]^{-1/2}\left(\frac{R_{\star}}{H_{\star}^{-1}}\right)^{-1}\times\\
&\quad&\times\left\{
        \begin{array}{ll}
            1.1 \times 10^{-11}\left(\frac{\rm {TeV}}{m_{\chi}}\right)\left(\frac{1 {\rm g}}{M_{\rm PBH}}\right)^{1/2} & \quad m_X<T_\text{PBH}, \\
            \\
           1.0 \times 10^{-19}\left(\frac{m_\chi}{10^{15}~{\rm GeV}}\right)\left(\frac{M_{\rm{PBH}}}{1 {\rm g}}\right)^{3/2} & \quad m_X>T_\text{PBH}.
        \end{array}
    \right.
\label{eq:eqbetaPBHFOPT}
\eea
For PBH formation from curvature perturbations,
\bea
\beta_{\rm PBH}&=&\frac{1}{g_\chi}\left(\frac{g_*(T_i)}{106.75}\right)^{1/4} \left(\frac{0.2}{\gamma}\right)^{1/2}\left(\frac{g_*(T_\text{PBH})}{106.75}\right)\times\nonumber\\
&\quad&\times\left\{
        \begin{array}{ll}
            8.0 \times 10^{-13}\left(\frac{\rm{TeV}}{m_\chi}\right)\left(\frac{1\,{\rm g}}{M_\text{PBH}}\right)^{1/2}& \quad m_X<T_\text{PBH}, \\
            \\
           7.1 \times 10^{-21}\left(\frac{m_\chi}{10^{15}\rm{GeV}}\right)\left(\frac{M_\text{PBH}}{1\,{\rm g}}\right)^{3/2} & \quad m_X>T_\text{PBH}.
        \end{array}
    \right.
\label{eq:eqbetaPBHzeta}
\eea

\begin{figure}[h]
  \centering    
  \includegraphics[width=0.55\textwidth]{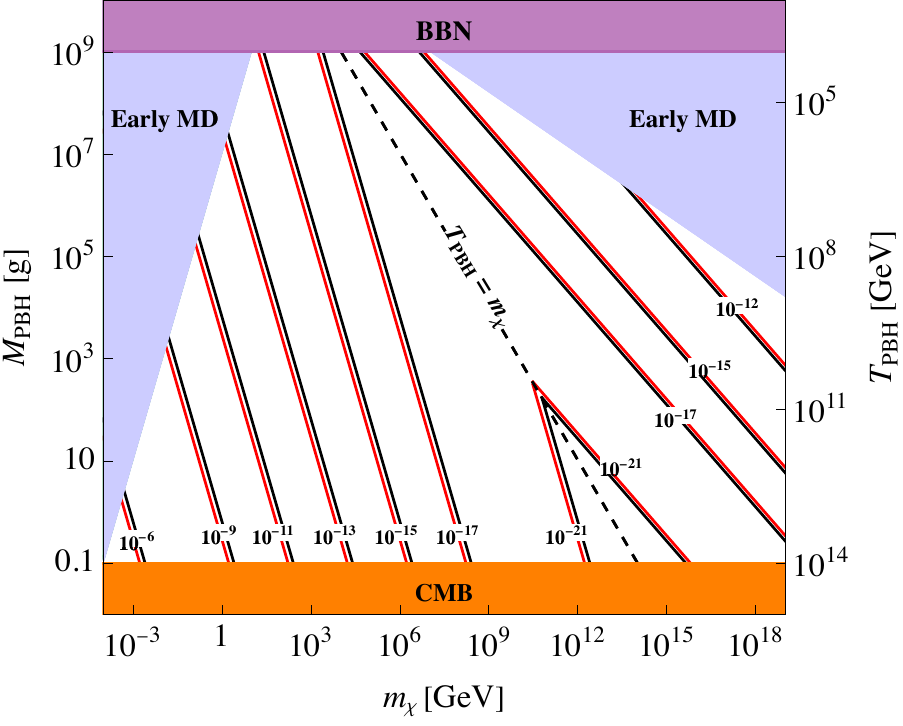}
  \caption{Contours of $\beta_{\rm PBH}$ on the plane of $M_{\rm PBH}$ vs. $m_{\chi}$ after requiring $\Omega_\chi h^2 = 0.12$ from FOPTs (red) assuming $R_{\star} = 1.5 H_{\star}^{-1}$, and from curvature perturbations (black). The black dashed curve show the PBH mass whose Hawking temperature is equal to the DM mass. Regions to the left of the black dashed curve means $m_\chi<T_{\rm PBH}$ and regions to the right of the black dashed curve means $m_\chi>T_{\rm PBH}$. For very light and very heavy DM masses, the relic abundance cannot be produced through Hawking radiation in a radiation-dominated Universe. These regions are shaded in blue for both mechanisms to indicate the early matter-dominated epoch. The upper bound on the PBH mass from BBN is shown in the purple shaded region. The lower bound on the PBH mass from cosmic microwave background (CMB) is shown in the orange shaded region.}
  \label{fig:betaDM1}
\end{figure}

\begin{figure}[h]
  \centering    
    \qquad\includegraphics[width=0.57 \textwidth]{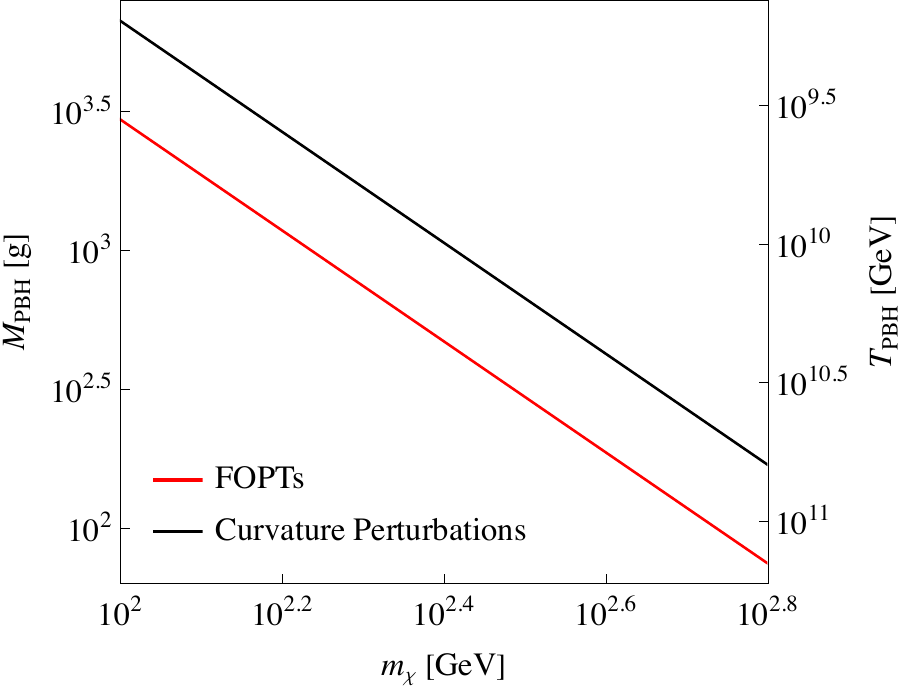}
  \caption{Same as Fig.~\ref{fig:betaDM1}, zoomed in on the contour $\beta_{\rm PBH} = 10^{-13}$ and requiring $\Omega_\chi h^2 = 0.12$ for the FOPT mechanism (red) and the curvature perturbation mechanism (black). We assume $R_{\star}=1.5 H_{\star}^{-1}$ for the red curve. }
  \label{fig:betaDM2}
\end{figure}

We are now in a position to describe the correlation between the DM relic density and PBHs. The results are plotted in Fig.~\ref{fig:betaDM1}. We show contours of $\beta_{\rm {PBH}}$ on the plane of $M_{\rm PBH}$ vs. $m_\chi$, after imposing $\Omega_\chi h^2 = 0.12$ \cite{Planck:2018vyg}. The contours corresponding to FOPT are shown in red, while the contours corresponding to scalar perturbations are shown in black. The contours for FOPT assume $R_{\star}/H_{\star}^{-1} = 1.5$. The regions shaded in blue indicate PBH-domination where the required $\beta_{\rm PBH}$ value is larger than the critical values in Eq.~\eqref{eq:betacFOPTRH} and Eq.~\eqref{eq:betacScalar}. We checked the PBH-domination regions of the two formation mechanisms are almost completely overlapping. The PBH-domination region on the top-left corner means the DM mass is too small such that the energy density of DM produced by a PBH is suppressed by the DM mass, while the PBH-domination region on the top-right corner means the suppression comes from the number of heavy DM particles that can be produced by a PBH. We focus on the PBH mass range of $0.1 {\rm g} \lesssim M_{\rm PBH} \lesssim 10^{9} {\rm g}$. The upper bound is coming from the requirement that PBHs should fully evaporate before the BBN to avoid any modification from the PBH Hawking radiation on successful BBN~\cite{Carr:2020gox}. The lower bound is set by the largest inflationary Hubble parameter $H_I / M_{\rm Pl} < 2.5\times10^{-5}$ that is allowed by the Planck cosmic microwave background (CMB) observation \cite{Planck:2018jri}. The purple and orange regions show the constraints from BBN and CMB, respectively. Along the black dotted line, the initial temperature of PBHs equals to the DM mass. This line divides the parameter space into two distinct regions: $m_\chi<T_{\rm PBH}$, corresponds to the left side, and, $m_\chi>T_{\rm PBH}$, corresponds to the right side. Since these two regions coincide along the black dotted line, the contours of initial abundance of PBHs on both sides should meet on this line, e.g, contours of $\beta_{\rm PBH}=10^{-21}$.

It is worth mentioning that light DM produced by PBH evaporation can be warm enough to erase small-scale structures via free-streaming. It is shown that DM with no non-gravitational interactions emitted by PBHs is not cold enough for $m_\chi\lesssim 1\,{\rm TeV}$ when $M_{\rm PBH}=10^9~{\rm g}$, and the lower limit on $m_\chi$ scales as $M_{\rm PBH}^{1/2}$ in order to avoid the free-streaming constraint~\cite{Gondolo:2020uqv}. Introducing non-gravitational interactions for DM can relax the bounds from structure formation. For example, kinetic equilibrium established with thermal contact with the SM sector after DM production can alleviate the free-streaming constraint without altering the existing DM yield (see~\cite{Bringmann:2006mu} for calculations of kinetic decoupling). DM self interactions can also relax this constraint~\cite{Bernal:2020kse}, but the number changing processes can change the reported DM yield in this study.

In Fig.~\ref{fig:betaDM2}, we enlarge the plot for a closer view of two curves from different formation mechanisms with fixed $\beta_{\rm PBH} = 10^{-13}$. Our main take-away lesson from Fig.~\ref{fig:betaDM2} is that the dependence of the DM yield on the PBH formation mechanism is from the ${T_i}/{M_{\rm PBH}}$ term in Eq. \eqref{eq:YRD2}. Since both mechanisms discussed in Section~\ref{sec:Formation} produce PBHs with the same scaling $M_{\rm PBH}\propto T^{-2}_i$, the black and red curves in Fig. \ref{fig:betaDM2} exhibit a high degree of similarity.

\subsection{High-frequency Gravitational Waves}

In this Section, we complete the maps $ \{M_{\rm PBH}, \beta_{\rm PBH}\} \, \rightarrow \{\alpha, \beta, T_\star, v_w \} \rightarrow \{h_c, f_{\rm GW}\}$  and $ \{M_{\rm PBH}, \beta_{\rm PBH}\} \, \rightarrow \{A_{\zeta}, k_p \} \rightarrow \{h_c, f_{\rm GW}\}$ from the PBH parameter space to the strain-frequency plane of GWs. Since both mechanisms generate peaked PBH mass spectra, we approximate the mass function to the monochromatic limit with the same $\{M_{\rm PBH}, \beta_{\rm PBH}\}$ values at the peak location. GW quantities appeared in this section are denoted with the subscript "sw" for sound wave contributions in the FOPT mechanism and with the subscript "$\zeta$" for the curvature perturbation mechanism. 

\begin{figure}[t]
  \centering
\includegraphics[width=0.5\textwidth]{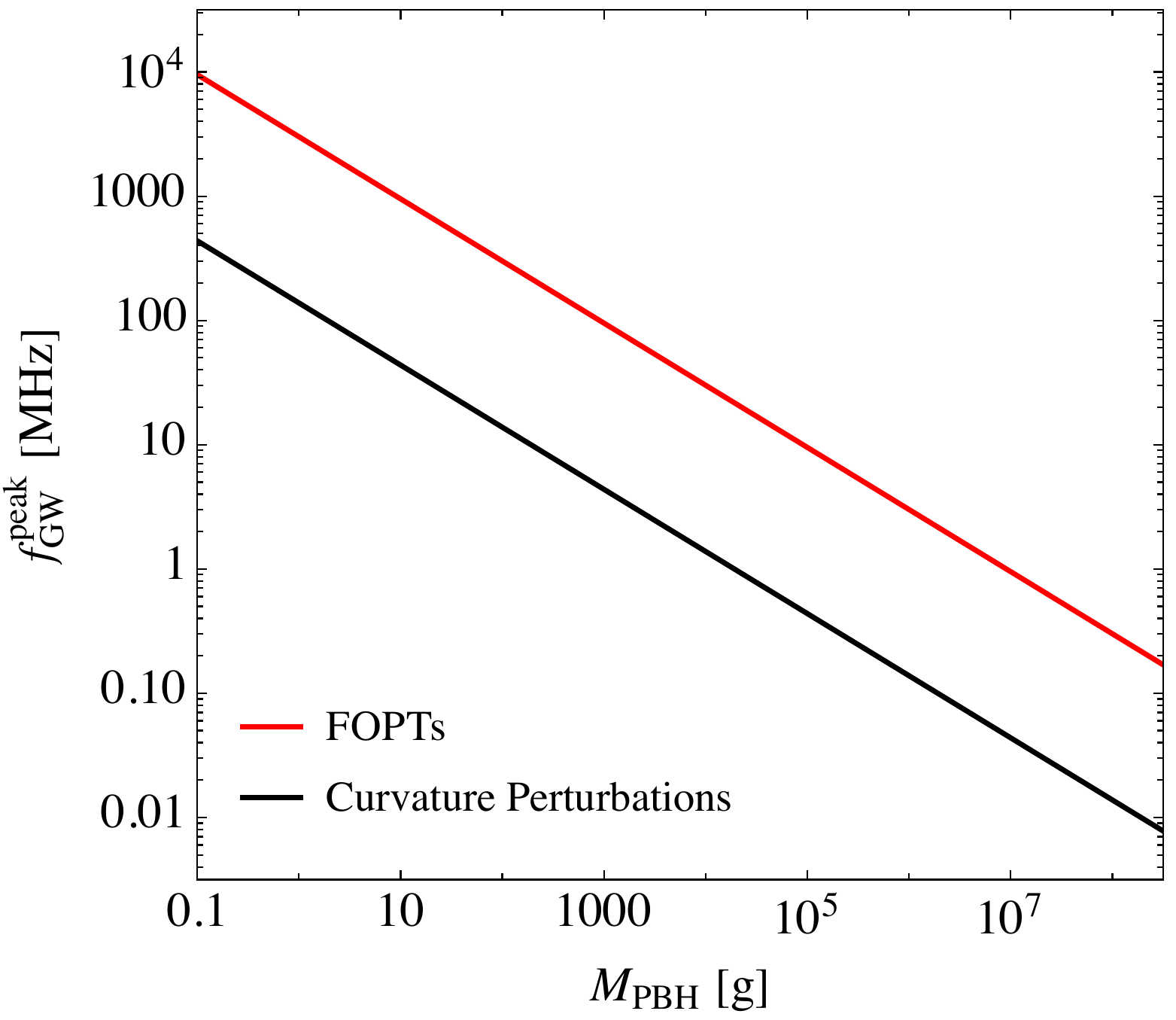}
  \caption{The peak frequency of ultra-high frequency GW spectra generated by sound waves during FOPTs (red) and induced by curvature perturbations (black), as a function of the PBH mass. The red curve is calculated with FOPT parameters $\alpha=0.8$, $\beta/H_\star=1.2$, and $v_w=0.5$. }
  \label{fig:fGWMPBH}
\end{figure}

We first calculate the peak frequencies in the two cases. Since our focus is on light PBHs that disappeared in the early Universe, the peak frequency is much higher than existing GW detectors. The peak GW frequencies in both mechanisms  have the same parametric dependence on the PBH mass $f^{\rm peak}_{\rm GW}\propto 1/\sqrt{M_{\rm PBH}}$. In the FOPT case, the peak frequency can be derived with Eqs. \eqref{eq:MPBHFOPT} and \eqref{eq:fGWpeakFOPT},
\bea
f_{\rm GW, sw}^{\rm peak}\simeq 30.1 \times \left(\frac{M_{\rm PBH}}{10^{4}~{\rm g}}\right)^{-\frac{1}{2}}{\rm MHz},
\label{eq:eq:fGWpeakokMPBHFOPT}
\eea
for the choice of FOPT parameters $v_w=0.5$ and $\beta/H_\star=1.2$. Adjusting $v_w$ and $\beta$ will only result in minor changes to the pre-factors and will not affect the overall scaling. On the other hand, the peak frequency of induced GWs is derived from Eqs. \eqref{eq:MPBHPzeta} and \eqref{eq:fGWpeakPzeta},\footnote{The peak location of the variance of density contrasts defined in Eq.~\eqref{eq:sigma0} differs from the peak location for the monochromatic power spectrum. In the numerical simulation, we find the variance peaks at $k\simeq0.7 k_p$. Therefore we included this factor when deriving $f_{{\rm GW},\zeta}^{\rm peak}$ from $M_{\rm PBH}$.}
\bea
f_{{\rm GW}, \zeta}^{\rm peak}\simeq 1.4 \times \left(\frac{M_{\rm PBH}}{10^{4}~{\rm g}}\right)^{-\frac{1}{2}}{\rm MHz}.
\label{eq:fGWpeakokMPBHPzeta}
\eea
The $f^{\rm peak}_{\rm GW}$ vs $M_{\rm PBH}$ relation is shown in Fig.~\ref{fig:fGWMPBH}. GWs produced during a FOPT (red) are typically at higher frequencies than GWs induced by curvature perturbations (black). Stated differently, the PBH mass from the FOPT is about two orders heavier than that from the curvature perturbation if the identical peak frequency in GWs is thought to originate from both formation mechanisms.

\begin{figure}[t]
  \centering
\includegraphics[width=0.45 \textwidth]{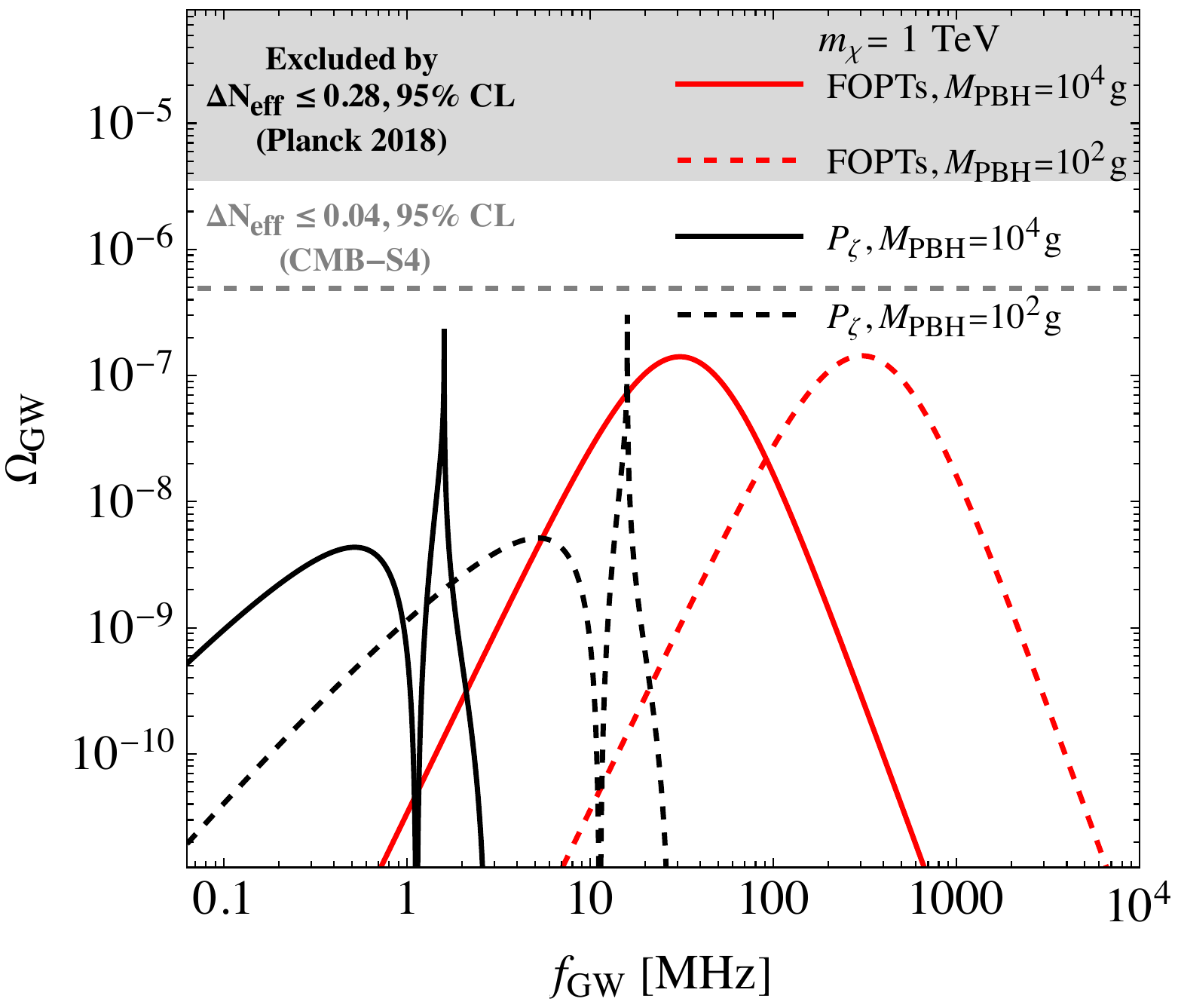}\quad
\includegraphics[width=0.45 \textwidth]{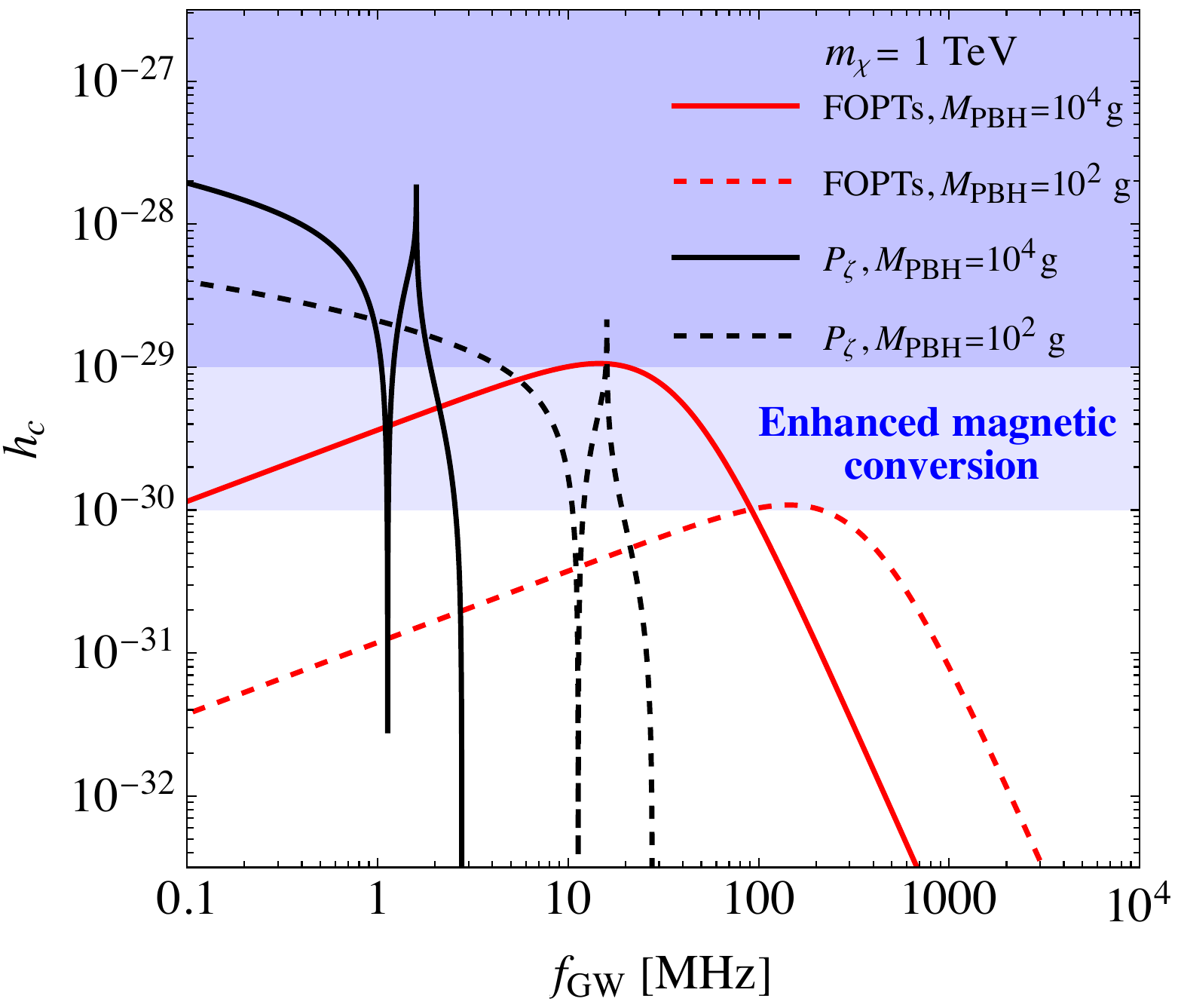}
  \caption{GW spectra for the case of producing $m_\chi=1~{\rm TeV}$ and $\Omega_\chi h^2=0.12$ from light PBHs of mass $M_{\rm PBH} = 10^4 {\rm g}$ (solid) and $M_{\rm PBH} = 10^2 {\rm g}$ (dashed). GWs produced by sound waves during FOPTs are shown in red color, and GWs induced by curvature perturbations are shown in black color. The FOPT parameters are $v_w=0.5$, $\alpha=0.8$, and $\kappa\simeq0.688$. {\it Left panel:} The GW density $\Omega_{\rm GW}$ as a function of the GW frequency. The current Planck $2\sigma$ constraint on GW contributions to $\Delta N_{\rm eff}$ is shown in the gray shaded region~\cite{Planck:2018vyg}. Future CMB-S4 $2\sigma$ sensitivity is shown with the gray dashed curve~\cite{CMB-S4:2016ple}. {\it Right panel:} The characteristic strain $h_c$ as a function of the GW frequency. The blue shaded regions show anticipated sensitivities of enhanced magnetic conversion detection using the inverse Gertsenshtein effect with conservative (dark blue) and optimistic (light blue) estimates.}
\label{fig:OmegaGWfGWBenchmark}
\end{figure}

In the next, we calculate the energy density of high frequency GWs for a benchmark DM mass $m_{\chi}=1~{\rm TeV}$ in Fig.~\ref{fig:OmegaGWfGWBenchmark} (left). Assuming the correct DM relic abundance $\Omega_\chi h^2 = 0.12$ is emitted by PBHs formed from FOPTs (red) and curvature perturbations (black), we obtain $\{M_{\rm PBH}, \beta_{\rm PBH}\}$ with Eq.~\eqref{eq:eqbetaPBHFOPT} and Eq.~\eqref{eq:eqbetaPBHzeta}. There is freedom in choosing either the mass or abundance of the PBHs for a given DM mass; we choose benchmarks $M_{\rm PBH} = 10^4 {\rm g}$ (solid) and $M_{\rm PBH} = 10^2 {\rm g}$ (dashed). The PBH abundance is $\beta_{\rm PBH}\simeq5.4\times10^{-15}$ for FOPTs and $\beta_{\rm PBH}\simeq8.2\times10^{-15}$ for curvature perturbations when $M_{\rm PBH} = 10^4 {\rm g}$. For a smaller PBH mass $M_{\rm PBH} = 10^2 {\rm g}$, the corresponding $\beta_{\rm PBH}$ values need to be exactly an order of magnitude larger for both mechanisms.

The next step in the procedure for obtaining $\Omega_{\rm GW}$ in the FOPT case is as follows. Given a point on the $\{M_{\rm PBH}, \beta_{\rm PBH}\}$ plane, we obtain the corresponding values of $T_\star$ from Eq. \eqref{eq:MPBHFOPT} assuming $R_{\star}/H^{-1}_{\star}=1.5$. We further fix the wall velocity to be $v_w = 0.5$ and obtain $\beta/H_\star$ by solving Eq. \eqref{eq:betapbhfinalfull} numerically. The phase transition temperature is $T_\star\simeq6.5\times10^{13}~{\rm GeV}$ and $T_\star\simeq6.5\times10^{14}~{\rm GeV}$ for $M_{\rm PBH} = 10^4 {\rm g}$ and $10^2 {\rm g}$ respectively. The value of $\beta/H_\star$ is found to be about $1.2$ and does not change appreciably as $\{M_{\rm PBH}, \beta_{\rm PBH}\}$ is varied. This is because the pocket distribution is double-exponentially sensitive to the nucleation rate, there a mild change in $\beta$ can give the correct PBH abundance. There are two additional inputs required:  $\alpha$ and $\kappa$, as can be seen from Eq.~\eqref{eq:OmegaGWFOPT}. We choose $\alpha=0.8$, which is found to be allowed for $v_w=0.5$ in Fig. 2 of \cite{Alves:2019igs}. We follow Appendix A of \cite{Guo:2021qcq} to calculate the $\kappa\simeq 0.688$ in the subsonic deflagration regime for our choice of $v_w$. The suppression factor in Eq.~\eqref{eq:FOPTGWUpsilon} is $\Upsilon\simeq0.71$ with our choice of parameters. This set therefore determines all the data in Eq.~\eqref{eq:OmegaGWFOPT}, which is used to obtain $\Omega_{\rm GW}$. The spectrum of GWs from sound waves has a smoother peak with the full spectrum shape determined by the shape function in the second line of Eq.~\eqref{eq:OmegaGWFOPT}.

Similarly, in the case of scalar perturbations, once a point on the $\{M_{\rm PBH}, \beta_{\rm PBH}\}$ plane is obtained for a given DM mass, Eq. \eqref{eq:MPBHPzeta} and \eqref{eq:betaPBHErfc} are used to obtain the peak location and amplitude of the monochromatic power spectrum. The primordial fluctuations appear at very small scales as $k_p\simeq8.9\times10^{20}~{\rm Mpc}^{-1}$ and $k_p\simeq8.9\times10^{21}~{\rm Mpc}^{-1}$ for $M_{\rm PBH}=10^{4}~{\rm g}$ and $M_{\rm PBH}=10^{2}~{\rm g}$ respectively. The amplitudes of monochromatic power spectra that generate heavier and lighter PBHs are $A_\zeta\simeq0.019$ and $A_\zeta\simeq0.020$. The benchmark power spectrum parameters are used to calculate induced GWs with Eq.~\eqref{eq:GWzeta}. The spectrum of induced GWs exhibits a resonant peak at $k=(2/\sqrt{3})k_p$ coming from the amplification when the GW source oscillates at a frequency equals to twice the frequency of the gravitational potential. In Fig.~\ref{fig:OmegaGWfGWBenchmark}, we show induced GWs using a delta-function power spectrum for simplicity. In the case where PBHs are generated by an extended power spectrum, the resulting GW spectrum shape is similar to the shape of the power spectrum, but without the presence of the divergent resonance. 

GWs generated in the early Universe contribute to the number of effective degrees of freedom as extra radiation. As a result, they are subject to the
cosmological constraint from BBN and CMB. This constraint is quantified as the effective number of neutrinos species after electron-positron annihilation, $N_{\rm 
 eff}$. Then any extra radiation, $\Delta \rho_{\rm rad}=\rho_{\rm GW}$ in our case, can be expressed in terms of extra neutrino specie, $\Delta N_{\rm eff}$ as
 \begin{equation}
   \Delta \rho_{\rm rad}=\frac{\pi^2}{30} \frac{7}{4} \left(\frac{4}{11}\right)^{\frac{4}{3}}\Delta N_{\rm eff} T_{\rm rad}^4.
 \end{equation}
The upper bound on $\Delta N_{\rm eff}$ sets an upper bound on $\Omega_{\rm GW}$. 
The Planck measurement result is $N_{\rm eff}=2.99 \pm 0.17$~\cite{Planck:2018vyg}. We require the GW contribution should not raise $N_{\rm eff}$ to be more than the $2\sigma$ upper bound set by Planck and shade the excluded regions in gray in the left panel of Fig.~\ref{fig:OmegaGWfGWBenchmark}. Future CMB-Stage 4 (CMB-S4) experiment is able to improve the sensitivity to $\sigma(N_{\rm eff})\simeq 0.02$-$0.03$~\cite{CMB-S4:2016ple}. We show future CMB-S4 $2\sigma$ sensitivity limit $\Delta N_{\rm eff}\leq 0.04$ with the gray dashed line.

While advancements in future cosmology observations continue to improve the sensitivity to primordial GWs, the positive detection of ultra-high frequency GWs remains a intriguing objective at the intersection of cosmology, particle physics and precision measurements. Various methods have been proposed to detect ultra-high frequency GWs with mechanical sensors~\cite{Harry:1996gh, Arvanitaki:2012cn, Aggarwal:2020umq, Goryachev:2014yra, Page:2020zbr,Goryachev:2021zzn}, interferometers~\cite{Chou_2017, Akutsu:2008qv}, conversion between GW and electromagnetic waves~\cite{Berlin:2021txa, Berlin:2022hfx, Herman:2022fau, Herman:2020wao, Domcke:2022rgu, Berlin:2023grv}, condensed matter systems~\cite{Ito:2019wcb} and radio telescopes~\cite{Domcke:2020yzq}. In the right panel of Fig.~\ref{fig:OmegaGWfGWBenchmark}, we show benchmark GW spectra on the $h_c$ vs $f_{\rm GW}$ plane. The strain strength is calculated from $\Omega_{\rm GW}$ with the definition
\bea
h_c=\frac{1}{f_{\rm GW}}\sqrt{\frac{3 \, H^2_0 \, \Omega_{\rm GW}}{4 \, \pi^2}},
\eea
where $H_0$ is the Hubble expansion rate today. In Fig.~\ref{fig:OmegaGWfGWBenchmark} (right), we also show expected future sensitivity reach using the inverse Gertsenshtein effect~\cite{gertsenshtein1962wave} to probe GWs with a Gaussian beam~\cite{Li:2003tv}, which improves the GW conversion rate to be only proportional to $h_c$. The dark blue region indicates the sensitivity is assumed to be $h_c\simeq 10^{-29}$, while the light blue region is assumed to have a more optimistic sensitivity of $h_c\simeq 10^{-30}$. We plot the sensitivity in a wide range of $f_{\rm GW}$ where resonant conversion requires the Gaussian beam and the single photon detector to work at the frequency of the target GWs. See also~\cite{Ringwald:2020ist} for discussions on future technological improvements to implement GW detection using Gaussian beams. In general, the future sensitivity to ultra-high frequency GWs has the potential to reach peak regions of GW spectra generated by PBH formation mechanisms.

\begin{figure}[t]
  \centering
\includegraphics[width=0.45\textwidth]{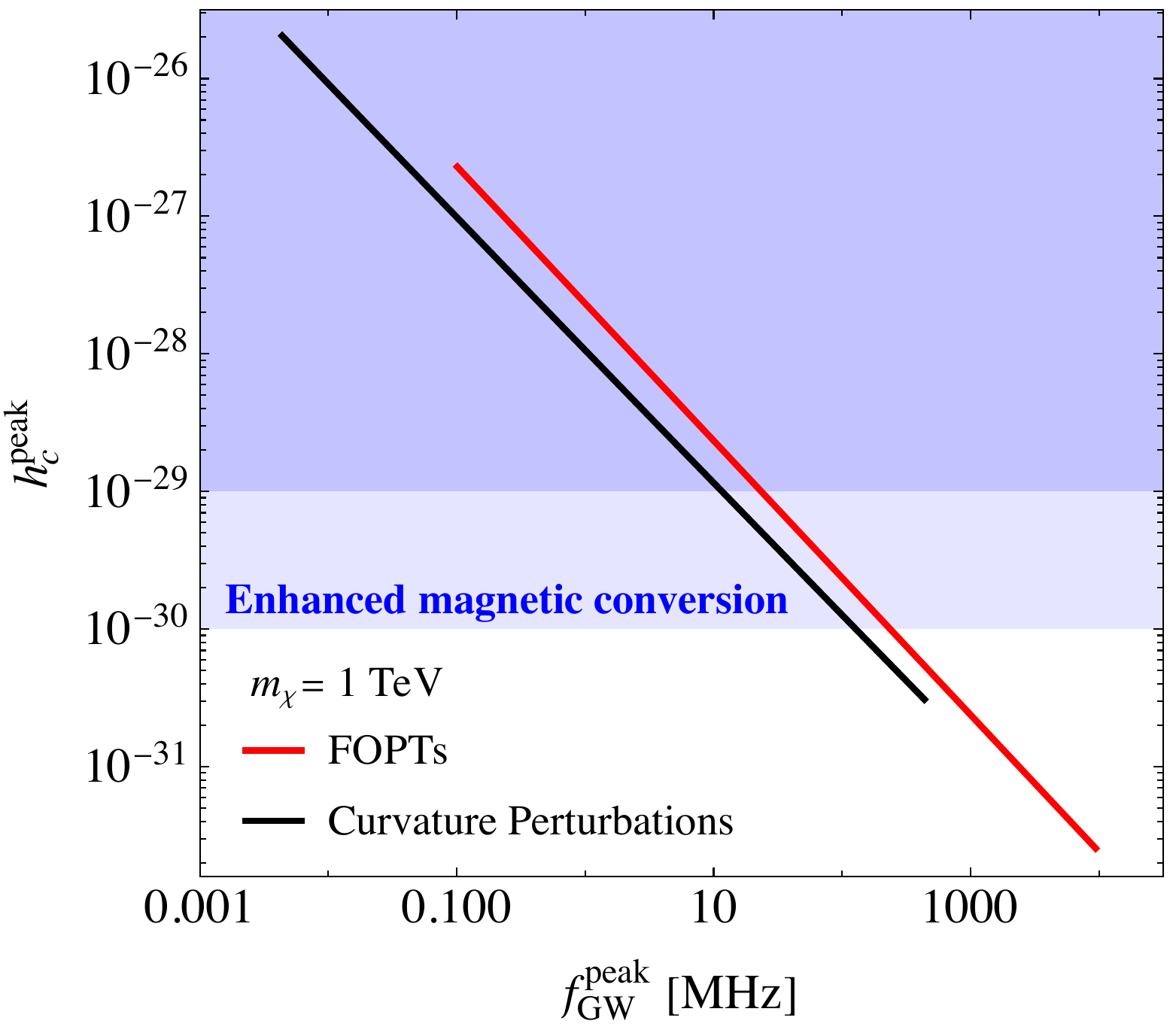}
\includegraphics[width=0.45\textwidth]{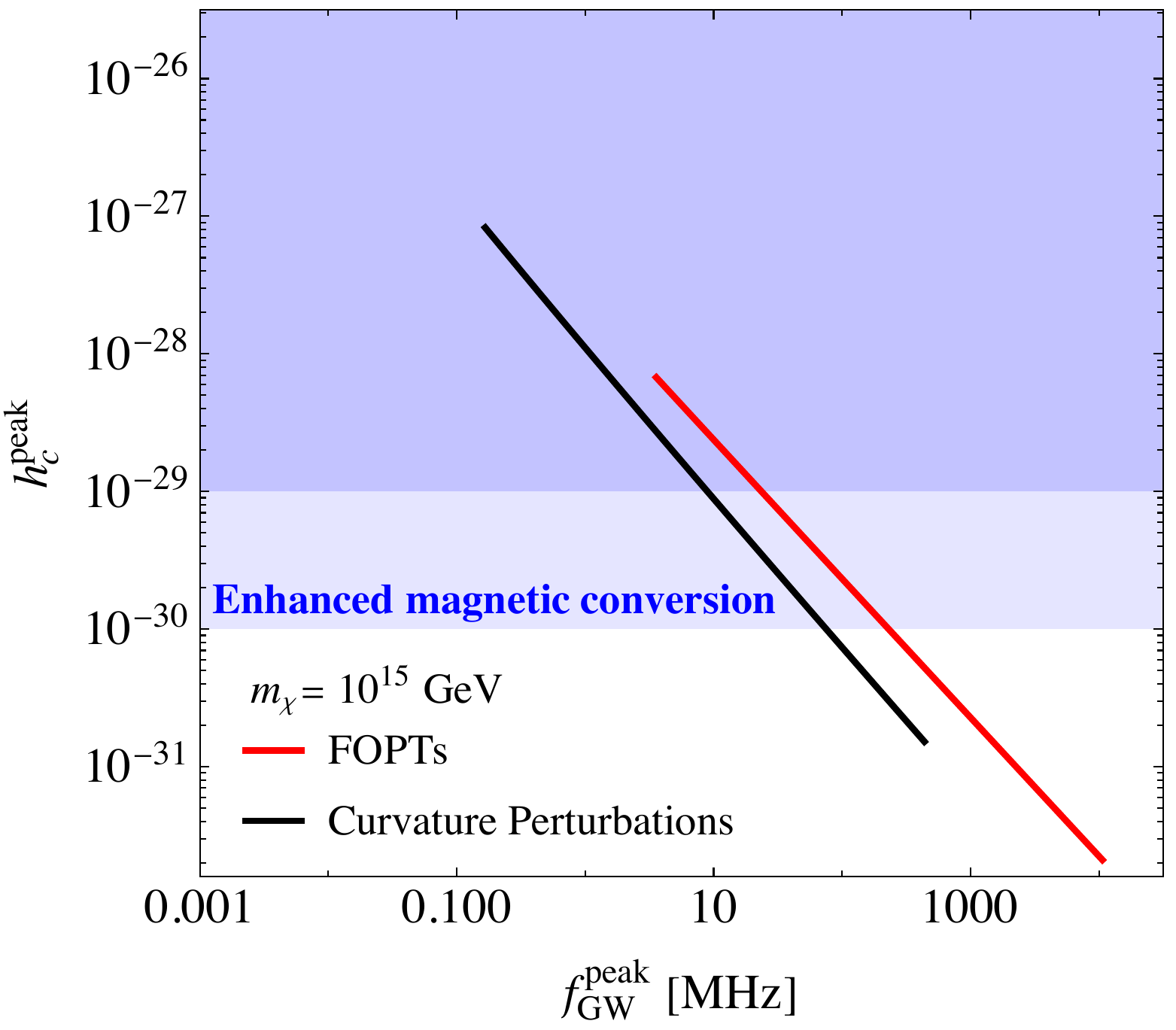}
  \caption{The peak strain strength $h^{\rm peak}_c$ and the peak frequency $f^{\rm peak}_{\rm GW}$ of GW spectra generated in the FOPT mechanism (red) and the curvature perturbation mechanism (black) that produce the DM relic abundance during a radiation-dominated era with $m_\chi = 1~{\rm TeV}$ (left) and $m_\chi = 10^{15}~{\rm GeV}$ (right), and with varying $M_{\rm PBH}$. The FOPT parameters are $v_w=0.5$, $\alpha=0.8$, and $\kappa\simeq0.688$. The induced GWs are calculated assuming the power spectrum has a finite width and there is no resonant amplification present for the induced GWs (see the text for explanation). The blue shaded regions show conservative (dark blue) and optimistic (light blue) sensitivity anticipations for future ultra-high frequency GW detection using the inverse Gertsenshtein effect.}
  \label{fig:hcfGWpeak}
\end{figure}

In Fig.~\ref{fig:hcfGWpeak}, we show the peak strength of GW signals for DM mass $m_\chi=1~{\rm TeV}$ (left) and $m_\chi=10^{15}~{\rm GeV}$ (right) produced by light PBHs, with the same color scheme used in Fig.~\ref{fig:OmegaGWfGWBenchmark} for two formation mechanisms. We choose PBH masses in the allowed region for the fixed DM mass in Fig.~\ref{fig:betaDM1}. For each $\{M_{\rm PBH}, m_\chi \}$, we calculate the peak frequency and peak strain values of GW spectra $\{f^{\rm peak}_{\rm GW} , h_{c}^{\rm peak}\}$, which are used to generate the curves. In the left panel, the correct DM relic abundance is always produced during the radiation-dominated era. Therefore, the left endpoints of curves are determined by the BBN constraint $M_{\rm PBH}<10^9~{\rm g}$ while the right endpoints are determined by the CMB constraint $M_{\rm PBH}>0.1~{\rm g}$. In the right panel, DM is heavier and its production from relatively heavy PBHs can only happen in the early matter-domination region shaded in the top-right corner of Fig.~\ref{fig:betaDM1}. For this reason, the left endpoints in the right panel are determined by the requirement $\beta_{\rm PBH}<\beta_{\rm PBH,crit}$, while the right endpoints are still set by the CMB constraint. We calculate the peak strain of FOPT curves (red) with Eq.~\eqref{eq:OmegaGWFOPT}. The peak strain of induced GWs (black) depends on the shape of the power spectrum. We formulated a conservative estimate by making the assumption that the power spectrum has a finite width, such that the resonant amplification has not been included in the GW spectrum. In this case, $\Omega_{\rm GW}^{\rm peak}\simeq A^2_\zeta \times \Omega_{{\rm rad},0}$ with $A_\zeta$ derived from the DM relic abundance. The peak strain $h_{c}^{\rm peak}$ is calculated accordingly. Both mechanisms predict ultra-high frequency GWs in the MHz-GHz range, offering a distinct signal strength benchmark for future searches.

\section{Conclusions}
\label{sec:conclusion}

In this paper we have studied the possible traces of light PBHs, which disappear before BBN, in the form of ultra-high frequency (MHz-GHz) GWs. More precisely, we have explored the signatures of PBH formation in the MHz-GHz frequency range of GW spectrum and their connections with the DM particles produced by Hawking evaporation of PBHs. The target frequency window of GWs is basically determined by the formation mechanism of PBHs.  

Assuming that DM particle has only gravitational interactions, and that PBHs would never dominate the energy density of the Universe,  the final abundance of DM is set by the mass and abundance of the PBHs and the mass of the DM particle. For a certain PBH formation mechanism, the part of the parameter space which gives rise to the observed relic abundance of DM today, leads to the correlated ultra-high frequency GWs and subsequently to the strain of the GWs as a function of frequency. 

Although there are a variety of mechanism for light PBH formation, in this study we focused on two specific formation mechanisms: PBH formation by curvature perturbations, and formation of PBHs from the collapse of particles trapped in the false vacuum of a FOPT.

For the canonical formation of PBHs by curvature perturbations, GWs are sourced by curvature perturbations with the second-order effects, and can be described by the amplitude of the power spectrum of curvature perturbations and the peak mode of the power
spectrum. 
For the formation of PBHs from FOPTs, the GWs, that are generated by the FOPT itself with the dominant contribution from sound waves in the fluid, are expressed in terms of the energy density released during phase transition normalized by
the radiation energy density, the inverse time scale of the phase transition, the temperature of the phase transition, and the bubble wall velocity. 

As we have showed, the dependence of the yield of DM on PBH formation mechanism is encoded in the ratio of the mass of the PBH to the temperature of the Universe at the formation time of the PBH. Since in both of the formation  mechanisms studied in this paper, the PBH mass follows the horizon mass at the formation time, therefore, they require almost the same initial abundance of PBHs to lead to the observed value of DM abundance today. 

After deriving the formation mechanism parameters by fixing the DM relic abundance, we find that GWs produced during a FOPT are typically at higher frequencies than GWs induced by curvature perturbations. For the same peak frequency, the PBH mass from the FOPT is about two orders heavier than that from the curvature perturbation. A more detailed computational treatment relying on solving Boltzmann equations for the benchmark FOPT temperature and nucleation rate from~\cite{Baker:2021sno} in the context of evaporating PBHs will be investigated in a future work.

As an example, we evaluated the energy density of high frequency GWs generated during the formation of PBHs with masses of $M_{\rm PBH} = 10^4 {\rm g}$ and $M_{\rm PBH} = 10^2 {\rm g}$ provided that the observed relic abundance of DM today is explained by DM particles with a mass of $m_{\chi}=1~{\rm TeV}$ emitted by PBHs.
For FOPT formation mechanism, the necessary initial abundances of PBHs of mass $M_{\rm PBH} = 10^4 {\rm g}$ and $M_{\rm PBH} = 10^2 {\rm g}$ are found to be equal to $\beta_{\rm PBH}\simeq5.4\times10^{-15}$ and $\beta_{\rm PBH}\simeq5.4\times10^{-14}$, which then determine the temperature of FOPTs to be equal to $T_\star\simeq6.5\times10^{13}~{\rm GeV}$ and $T_\star\simeq6.5\times10^{14}~{\rm GeV}$ respectively. GW spectra generated by sound waves during
these FOPTs peak at $30\,{\rm MHz}$ and $300\,{\rm MHz}$ respectively. Since the pocket distribution is double-exponentially
sensitive to the nucleation rate, variations in the mass and initial abundance of PBHs only change the required nucleation rate slightly.

In the case of scalar perturbations, PBHs with masses equal to $M_{\rm PBH} = 10^4 {\rm g}$ and $M_{\rm PBH} = 10^2 {\rm g}$ with initial abundances of 
$\beta_{\rm PBH}\simeq8.2\times10^{-15}$ and $\beta_{\rm PBH}\simeq8.2\times10^{-14}$, respectively, can explain the observed abundance of DM today.
For $M_{\rm PBH}=10^{4}~{\rm g}$, the primordial fluctuations appear at $k_p\simeq8.9\times10^{20}~{\rm Mpc}^{-1}$ which corresponds to GW signal with a peak frequency of $1\,{\rm MHz}$. For $M_{\rm PBH}=10^{2}~{\rm g}$, these parameters are given as $k_p\simeq8.9\times10^{21}~{\rm Mpc}^{-1}$ and $10\,{\rm MHz}$, respectively.

Although the projected sensitivity of the future CMB-S4 experiment in probing stochastic GW signals from early Universe is almost one order of magnitude better than Planck, it is still a few times above the predicted signals by the PBH formation mechanisms studied in this paper. On the other hand, the expected future
sensitivity reach of the enhanced magnetic conversion detection by utilizing the inverse Gertsenshtein effect can probe the peaks of the predicted GW signals in this study. 

As we demonstrate here, PBH formation via FOPTs and curvature perturbations, both can give rise to production of ultra-high frequency (MHz-GHz) GWs. These GW signals, with distinct strength and frequency spectrum, could potentially fall into the reach of future searches and can be used to differentiate between different PBH formation mechanisms. Possible correlations of new physics with PBH formation mechanism, e.g. DM production through the Hawking evaporation of PBHs, is also coded in the associated GW signals, which make the MHz-GHz GW searches a promising frontier to pursue.

\acknowledgments
We would like to thank Huai-Ke Guo for useful discussions. The work of B.S.E is supported in part by DOE Grant DE-SC0022021.
The work of K.S. and T.X. is supported in part by DOE Grant desc0009956.

\bibliographystyle{JHEP}
\bibliography{article}
\end{document}